\documentclass[essd, manuscript]{copernicus}

\usepackage{tabularx}
\usepackage{float}
\usepackage{colortbl}
\usepackage{xcolor}

\nolinenumbers

\begin{document}

\title{OpenSWI: A Massive-Scale Benchmark Dataset for Surface Wave Dispersion Curve Inversion}

%
%
\Author[1,2]{Feng}{Liu}
\Author[2,3]{Sijie}{Zhao}
\Author[2]{Xinyu}{Gu}
\Author[2]{Fenghua}{Ling}
\Author[2]{Peiqin}{Zhuang}
\Author[4][yxli2024@cdut.edu.cn]{Yaxing}{Li} 
\Author[2][surui@pjlab.org.cn]{Rui}{Su} 
\Author[5]{Lihua}{Fang}
\Author[5]{Lianqing}{Zhou}
\Author[4]{Jianping}{Huang}
\Author[2]{Lei}{Bai}

\affil[1]{School of Electronic Information and Electrical Engineering, Shanghai Jiao Tong University, Shanghai 200240, China}
\affil[2]{Shanghai Artificial Intelligence Laboratory, Shanghai 200232, China}
\affil[3]{School of Geography and Ocean Science, Nanjing University, Nanjing 210023, China}
\affil[4]{Chengdu University of Technology, Chengdu 610059, China}
\affil[5]{Institute of Earthquake Forecasting, China Earthquake Administration, Beijing 100036, China}

\runningtitle{OpenSWI}

\runningauthor{Liu et al.}

\received{}
\pubdiscuss{} 
\revised{}
\accepted{}
\published{}


\firstpage{1}

\maketitle
    
%
%
\begin{abstract}
    Surface wave dispersion curve inversion plays a critical role in both shallow resource exploration and deep geological studies, yet it remains hindered by sensitivity to initial models, susceptibility to local minima, and low computational efficiency. Recently, data-driven deep learning methods, inspired by their success in computer vision and natural language processing, have shown promising potential to overcome these challenges. However, the lack of large-scale and diverse benchmark datasets remains a major obstacle to the development and evaluation of such methods. To address this gap, we introduce \textbf{OpenSWI}, a comprehensive benchmark dataset generated through the surface wave inversion dataset preparation (SWIDP) pipeline. OpenSWI comprises two synthetic datasets tailored to different research scales and application scenarios, namely \textbf{OpenSWI-shallow} and \textbf{OpenSWI-deep}, as well as an AI-ready real-world dataset for generalization evaluation, \textbf{OpenSWI-real}. OpenSWI-shallow is derived from the 2-D geological model dataset OpenFWI, containing over 22 million 1-D velocity profiles paired with their fundamental-mode phase and group velocity dispersion curves, spanning a broad spectrum of shallow geological structures (e.g., flat layers, faults, folds, and realistic stratigraphy). OpenSWI-deep is built from 14 global and regional 3-D geological models, comprising approximately 1.26 million high-fidelity 1-D velocity-dispersion data pairs for deep earth studies. OpenSWI-real, compiled from open-source projects, contains two sets of observed dispersion curves and their corresponding 1-D reference models, serving as a benchmark for evaluating the generalization of deep learning models. To demonstrate the utility of OpenSWI, we trained deep learning models on OpenSWI-shallow and OpenSWI-deep, and evaluated them on OpenSWI-real. The results show strong agreement between the predicted and reference velocity models, confirming the diversity and representativeness of the OpenSWI dataset. To facilitate the advancement of intelligent surface wave dispersion curve inversion techniques, we release the SWIDP toolbox, the OpenSWI datasets, trained deep learning models, and other examples, aiming to provide comprehensive support and open resources for the research community.
\end{abstract}

\copyrightstatement{} 

%
%
\introduction  
    Surface wave dispersion curve inversion is a fundamental geophysical technique for reconstructing subsurface shear wave velocity profiles by fitting theoretical dispersion curves to measured data \citep{xia_1999_Estimation, shapiro_2004_Emergence, wathelet_2004_Surfacewave}. It is widely applied in shallow engineering surveys, including site response and microzonation studies \citep{park_1999_Multichannel, socco_2004_Surfacewave, foti_2014_Surface}, as well as in studies of lithospheric structure and evolution at greater depths \citep{shapiro_2002_MonteCarlo, shapiro_2004_Emergence, yang_2008_Characteristics}. In shallow subsurface investigations, this technique is valuable for identifying complex geological features such as weathering layers and overburden, while at greater depths, it provides critical insights into tectonic evolution \citep{reid_2025_Ambient}. Despite its widespread applicability, traditional inversion methods are heavily dependent on initial models and nonlinear optimization, which results in high computational costs and susceptibility to local minima \citep{shapiro_2002_MonteCarlo, wathelet_2004_Surfacewave, chen_2025_Why}. These limitations hinder their applicability to large-scale, high-resolution imaging tasks.

    In recent years, rapidly developed deep learning methods have revolutionized the process of surface wave dispersion curve inversion. These data-driven approaches leverage deep neural networks, such as fully connected networks (FNNs), convolutional neural networks (CNNs), and Transformer networks, to learn the mapping between dispersion curves and subsurface shear wave velocity profiles \citep{hu_2020_Using, yablokov_2021_Artificial, wang_2022_Deep, cai_2022_SemiSupervised, huang_2024_JointNet, liu_2025_DispFormer, jiang_2025_OneFitAll}. By effectively eliminating reliance on initial models and iterative optimization, these methods significantly improve inversion efficiency and performance \citep{chen_2025_Why}. Once trained, the models can rapidly invert large-scale datasets in seconds, making them well-suited for real-time applications, such as field deployment and imaging. However, their performance and generalization ability are strongly influenced by the quality and diversity of the training data \citep{luo_2022_Constructing}. Previous research has demonstrated that large-scale, diverse datasets substantially enhance deep model performance, particularly in scenarios with no labeled data (zero-shot learning) or limited labeled data that requires fine-tuning (few-shot learning) \citep{luo_2022_Constructing, liu_2025_DispFormer}. Therefore, the development of dispersion curve datasets that encompass representative geological features, multi-scale structures, and sufficient sample sizes is crucial for advancing intelligent inversion methods.

    Despite the importance of diverse datasets for deep learning methods, the construction of benchmark datasets specifically for surface wave dispersion curve inversion remains limited. In contrast, other areas of seismic research have seen the successful creation of large-scale datasets. For instance, in seismic monitoring, datasets like STEAD \citep{mousavi_2019_STanford} and INSTANCE \citep{michelini_2021_INSTANCE} contain millions of waveform data traces. Similarly, full-waveform inversion efforts have led to the creation of model collections such as OpenFWI \citep{deng_2021_OpenFWI} and EFWI \citep{feng_2023_EFWI}, each comprising hundreds of thousands of geological velocity models. Seismic exploration has also benefited from the development of standardized workflows, leading to benchmark datasets like cigFacies \citep{gao_2025_CigFacies} and cigChannels \citep{wang_2025_CigChannel}. However, in the specific domain of surface wave dispersion curve inversion, there is still a significant lack of representative, well-structured, and publicly accessible datasets. One of the main challenges lies in the necessity of paired dispersion curves and velocity profiles to generate high-quality training samples. Actual observational data are often proprietary and not available to most of the researchers \cite{merrifield_2022_Synthetic}. Moreover, observed dispersion curves are often compromised by limitations in observation conditions and subjective picking, resulting in issues such as noise contamination and data missing \citep{socco_2004_Surfacewave, bensen_2007_Processing}. Additionally, the non-uniqueness of the corresponding velocity profiles further complicates the development of supervised models \citep{foti_2009_Nonuniqueness}, making it more difficult to train deep learning algorithms effectively.

    To address those challenges, synthetic surface wave dispersion curve data has emerged as a feasible alternative. Synthetic data, generated through a series of forward modeling processes, can effectively simulate field-observed dispersion curves. Since the corresponding velocity profiles are known in the simulation, this method naturally avoids pairing errors. Deep neural networks trained on synthetic data have demonstrated good applicability and inversion performance in shallow subsurface geological exploration \citep{cao_2020_Nearrealtime, aleardi_2021_Hybrid, yablokov_2021_Artificial, yablokov_2023_Uncertainty, gan_2024_Deep} and deep structural imaging \citep{hu_2020_Using, wang_2022_Deep, huang_2024_JointNet, jiang_2025_OneFitAll, liu_2025_DispFormer}. However, existing publicly available datasets are still largely limited to specific geological features or particular regions, lacking sufficient geological diversity and regional coverage. Given the complexity of shallow geology and the regional variability of deep structures, constructing a synthetic dataset with greater geological complexity, broader coverage, and larger sample sizes is essential for improving the generalization ability and practical applicability of models.

\begin{figure*}[!ht]
        \centering
        \includegraphics[width=1.0\textwidth]{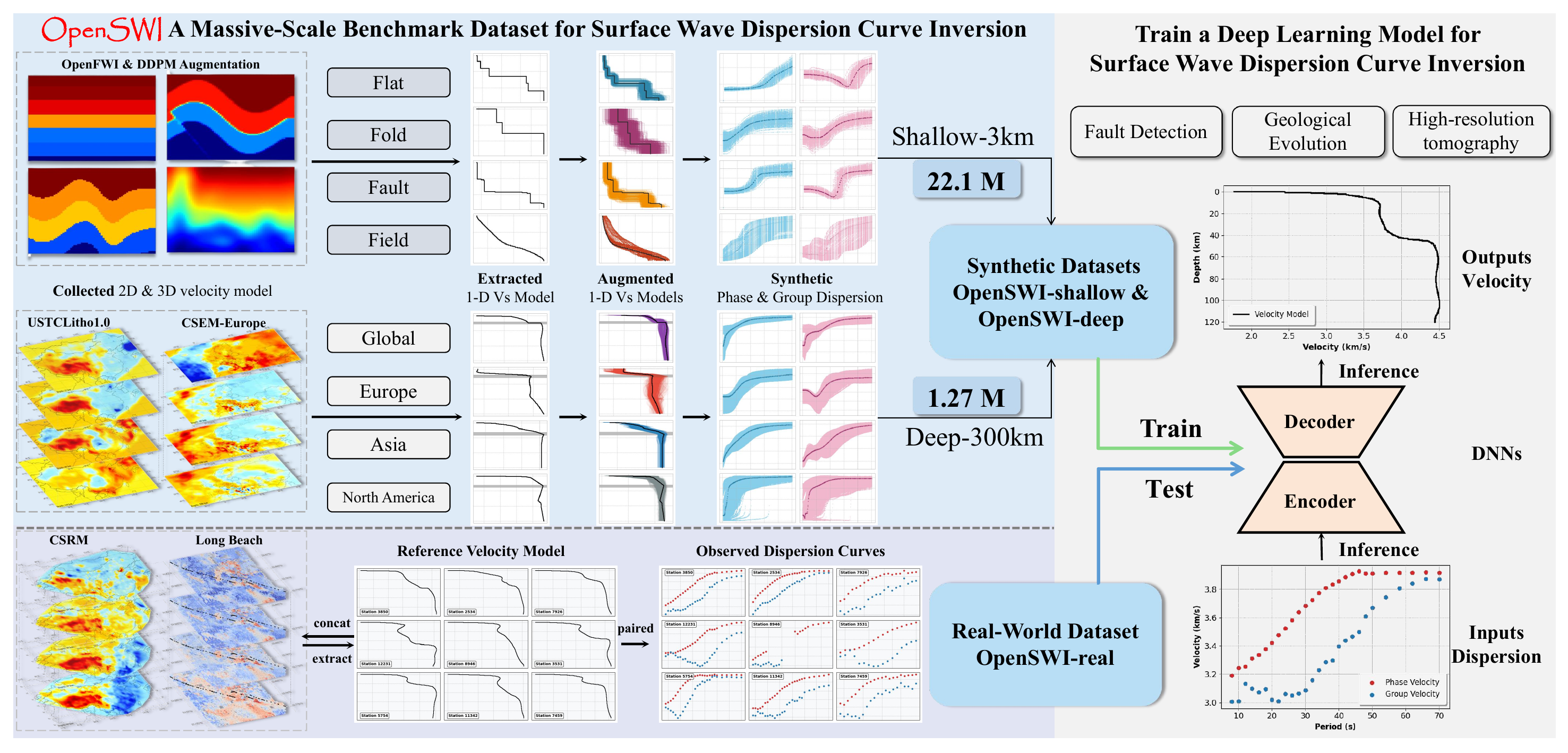}
        \caption{Overview of the workflow for constructing the OpenSWI benchmark datasets, and their application in deep learning-based surface wave dispersion curve inversion. The workflow starts with the collection and quality control of raw data, followed by the extraction and augmentation of 1-D velocity profiles, and the simulation of dispersion curves to generate two synthetic datasets, OpenSWI-shallow and OpenSWI-deep, tailored for different research scales and application scenarios (blue box). To evaluate the generalization capability of deep learning models, a real-world dataset, OpenSWI-real, is also curated (purple box). Finally, a simple deep learning model, trained on the benchmark datasets, is applied to real observational data, as depicted in the white box on the right.}
        \label{fig1:fwi_types}
    \end{figure*}

    In this paper, we introduce OpenSWI, a comprehensive benchmark dataset designed for surface wave dispersion curve inversion, developed through the dataset construction workflow SWIDP (Figure~\ref{fig1:fwi_types}). OpenSWI includes two synthetic benchmark datasets, OpenSWI-shallow and OpenSWI-deep, each tailored to different research scales and application scenarios, as well as an AI-ready real-world dataset, OpenSWI-real, specifically for evaluating model generalization. The OpenSWI-shallow dataset, built upon the publicly available 2-D geological model dataset OpenFWI, incorporates a broad range of geological features, such as flat layers, faults, folds, and actual geological structures, containing approximately 22 million 1-D velocity profiles paired with their corresponding fundamental-mode surface wave dispersion curves. This makes it the largest and most geologically diverse dataset available for shallow subsurface studies. To further enhance structural diversity and sample variability, SWIDP integrates a Diffusion Probabilistic Model (DDPM), which learns the distribution of 2-D geological models and allows the continuous generation of more varied shallow subsurface data. The OpenSWI-deep dataset, generated by collecting, curating, and integrating 14 global and regional 3-D geological models, consists of approximately 1.26 million high-fidelity 1-D dispersion data samples, providing a large-scale benchmark for deep subsurface imaging tasks. OpenSWI-real, derived from two publicly available observational datasets and their reference velocity models, is directly applicable for performance testing and generalization validation of deep learning models in real-world applications. To evaluate the practical utility of these datasets, we trained two Transformer-based models using OpenSWI-shallow and OpenSWI-deep, then validated them on OpenSWI-real. Experimental results show that the inversion results of the trained models on real-world data are highly consistent with reference models, confirming the effectiveness and representativeness of the OpenSWI datasets for real-world applications. All datasets, along with the associated toolchain (including profile extraction, forward modeling and training examples), have been fully open-sourced, offering a reusable, high-quality benchmark platform for advancing future research in intelligent surface wave dispersion curve inversion.

%
%
\section{Construction of the Large-scale OpenSWI Benchmark Datasets}

%
%
\subsection{Integrated Workflow for Dataset Construction}

    We present an integrated workflow for constructing large-scale benchmark datasets for surface wave dispersion curve inversion. The workflow is designed to ensure geological diversity and realism of the data sources, employ modular and fully automation processing, and ensure high accuracy and computational efficiency in forward modeling. It encompasses all major stages, from the collection and standardization of raw geological models, through quality control and parameterization, to the simulation of fundamental-mode dispersion curves, providing a reproducible pipeline for large-scale dataset generation.
    
\subsubsection{Collection and Quality Control of Geological Models}
\label{sec:collection_and_QC}

    The first step in constructing a high-quality dataset for dispersion curve inversion is the collection of representative velocity models from diverse geological settings. These velocity models were primarily obtained from open-access geological databases and previously published studies, such as OpenFWI datasets \citep{deng_2021_OpenFWI}—which contain 2-D geological models covering various sedimentary and tectonic settings—and LITHO1.0 geological models \citep{pasyanos_2014_LITHO10}, providing lithospheric-scale structural information. Table~\ref{table1:data_source} summarizes the original data sources employed in this study. These rigorously curated and geologically validated models form a reliable foundation for constructing the OpenSWI datasets.
    However, because the raw velocity models originated from different research groups and projects, they exhibited considerable variability in several aspects, such as data characteristics (e.g., depth range, spatial resolution), parameter types (e.g., S-wave velocity (\(v_s\)), P-wave velocity (\(v_p\)), or combined shear-wave velocities in both horizontal and vertical directions (\(v_{sv}\) and \(v_{sh}\))), and storage formats (e.g., \texttt{.npz}, \texttt{.txt}, or \texttt{.nc}). As a result, a unified quality control and standardization process was applied to ensure data consistency and representativeness before their use. The quality control procedures included the following steps:
    \begin{enumerate}
        \item \textbf{Data correction and cleaning:} Models containing missing or abnormal values (e.g., \texttt{zero} or \texttt{NaN} values in some geological models) were corrected through interpolation or single-point removal to ensure data completeness.
        \item \textbf{Parameter conversion:} For models that provided only $v_p$, the corresponding $v_s$ were estimated using the empirical relationships proposed by \citet{brocher_2005_Empirical}. In cases where models included $v_{sv}$ and $v_{sh}$, an equivalent $v_s$ was derived using the geometric mean.
        \item \textbf{Plausibility verification:} Geological structures within the models were systematically examined to remove anomalies inconsistent with geological principles or unsuitable for forward modeling.
    \end{enumerate}
        
    These quality control measures substantially improved the accuracy and applicability of the geological models, thereby providing a robust and standardized data foundation for dispersion curve forward modeling and subsequent machine learning model training.

\begin{table}[!ht]
\renewcommand{\arraystretch}{1.1}

\caption{Original data sources used in constructing the OpenSWI datasets, summarizing dataset categories (e.g., OpenSWI-shallow, OpenSWI-deep, and OpenSWI-real), references, primary geological settings (e.g., Flat, Flat-Fault, Fold, Fold-Fault, and Field) or geographic coverage (e.g., global, China, Europe, the United States), recorded velocity parameters (e.g., P-wave velocity $v_p$, S-wave velocity $v_s$, combined shear-horizontal velocity $v_{sh}$, and shear-vertical velocity $v_{sv}$), as well as the size of the raw data, expressed as $N$~velocity~profiles~$\times$~$M$~model~variables~$\times$~2-D~velocity~model~shape (for OpenSWI-shallow) or $L$~layers (for OpenSWI-deep and OpenSWI-real).}

\begin{tabular}{c|ccccc}
\hline
Group                                                                       & Reference                                                                                                                       & Datasets                                       & \begin{tabular}[c]{@{}c@{}}Geological Feature\\ /Cover Region\end{tabular}              & Model Variable                                      & Model Size                                   \\ \hline
                                                                            &                                                                                                                                 & \cellcolor[HTML]{FFFFFF}OpenFWI-FlatVelA       & \cellcolor[HTML]{FFFFFF}Flat                                                            & \cellcolor[HTML]{FFFFFF}$v_p$                       & \cellcolor[HTML]{FFFFFF}30,000 $\times$ 1 $\times$ 70 $\times$ 70 \\
                                                                            &                                                                                                                                 & \cellcolor[HTML]{EFEFEF}OpenFWI-Flat-FaultA    & \cellcolor[HTML]{EFEFEF}Flat + Fault                                                    & \cellcolor[HTML]{EFEFEF}$v_p$                       & \cellcolor[HTML]{EFEFEF}54,000 $\times$ 1 $\times$ 70 $\times$ 70 \\
                                                                            &                                                                                                                                 & \cellcolor[HTML]{FFFFFF}OpenFWI-CurveVel       & \cellcolor[HTML]{FFFFFF}Fold                                                            & \cellcolor[HTML]{FFFFFF}$v_p$                       & \cellcolor[HTML]{FFFFFF}30,000 $\times$ 1 $\times$ 70 $\times$ 70 \\
                                                                            &                                                                                                                                 & \cellcolor[HTML]{EFEFEF}OpenFWI-Fold-Fault     & \cellcolor[HTML]{EFEFEF}Fold + Fault                                                    & \cellcolor[HTML]{EFEFEF}$v_p$                       & \cellcolor[HTML]{EFEFEF}54,000 $\times$ 1 $\times$ 70 $\times$ 70 \\
\multirow{-5}{*}{\begin{tabular}[c]{@{}c@{}}OpenSWI\\ shallow\end{tabular}} & \multirow{-5}{*}{\citet{deng_2021_OpenFWI}}                                                                                     & \cellcolor[HTML]{FFFFFF}OpenFWI-StyleA         & \cellcolor[HTML]{FFFFFF}Field                                                           & \cellcolor[HTML]{FFFFFF}$v_p$                       & \cellcolor[HTML]{FFFFFF}67,000 $\times$ 1 $\times$ 70 $\times$ 70 \\ \hline
                                                                            & \cellcolor[HTML]{EFEFEF}\citet{pasyanos_2014_LITHO10}                                                                           & \cellcolor[HTML]{EFEFEF}LITHO1.0               & \cellcolor[HTML]{EFEFEF}Global                                                          & \cellcolor[HTML]{EFEFEF}$depth$, $v_s$              & \cellcolor[HTML]{EFEFEF}40,962 $\times$ 2 $\times$ 96      \\
                                                                            & \cellcolor[HTML]{FFFFFF}\citet{xin_2019_Highresolution}                                                                         & \cellcolor[HTML]{FFFFFF}USTClitho1.0           & \cellcolor[HTML]{FFFFFF}China                                                           & \cellcolor[HTML]{FFFFFF}$depth$, $v_s$              & \cellcolor[HTML]{FFFFFF}9,125 $\times$ 2 $\times$ 12       \\
                                                                            & \cellcolor[HTML]{EFEFEF}\citet{shen_2013_3D}                                                                                    & \cellcolor[HTML]{EFEFEF}Central-and-Western US & \cellcolor[HTML]{EFEFEF}USA                                                             & \cellcolor[HTML]{EFEFEF}$depth$, $v_s$              & \cellcolor[HTML]{EFEFEF}6,803 $\times$ 2 $\times$ 72       \\
                                                                            & \cellcolor[HTML]{FFFFFF}\citet{shen_2016_Seismic}                                                                               & \cellcolor[HTML]{FFFFFF}Continental China      & \cellcolor[HTML]{FFFFFF}China                                                           & \cellcolor[HTML]{FFFFFF}$depth$, $v_s$              & \cellcolor[HTML]{FFFFFF}4,516 $\times$ 2 $\times$ 400      \\
                                                                            & \cellcolor[HTML]{EFEFEF}\citet{xie_2018_3D}                                                                                     & \cellcolor[HTML]{EFEFEF}US Upper-Mantle        & \cellcolor[HTML]{EFEFEF}USA                                                             & \cellcolor[HTML]{EFEFEF}$depth$, $v_s$              & \cellcolor[HTML]{EFEFEF}3,678 $\times$ 2 $\times$ 600      \\
                                                                            & \cellcolor[HTML]{FFFFFF}\citet{lu_2018_Highresolution}                                                                          & \cellcolor[HTML]{FFFFFF}EUcrust                & \cellcolor[HTML]{FFFFFF}European                                                        & \cellcolor[HTML]{FFFFFF}$depth$, $v_s$              & \cellcolor[HTML]{FFFFFF}43,520 $\times$ 2 $\times$ 80      \\
                                                                            & \cellcolor[HTML]{EFEFEF}\citet{berg_2020_Shear}                                                                                 & \cellcolor[HTML]{EFEFEF}Alaska                 & \cellcolor[HTML]{EFEFEF}Alaska                                                          & \cellcolor[HTML]{EFEFEF}$depth$, $v_s$              & \cellcolor[HTML]{EFEFEF}19,408 $\times$ 2 $\times$ 156     \\
                                                                            & \cellcolor[HTML]{FFFFFF}\begin{tabular}[c]{@{}c@{}}\citet{sabuncu_2017_3D}\\ \citet{blom_2020_Seismic}\end{tabular}             & \cellcolor[HTML]{FFFFFF}CSEM-Europe            & \cellcolor[HTML]{FFFFFF}European                                                        & \cellcolor[HTML]{FFFFFF}$depth$, $v_{sh}$, $v_{sv}$ & \cellcolor[HTML]{FFFFFF}21,931 $\times$ 3 $\times$ 61      \\
                                                                            & \cellcolor[HTML]{EFEFEF}\citet{blom_2020_Seismic}                                                                               & \cellcolor[HTML]{EFEFEF}CSEM-Eastmed           & \cellcolor[HTML]{EFEFEF}\begin{tabular}[c]{@{}c@{}}Eastern\\ Mediterranean\end{tabular} & \cellcolor[HTML]{EFEFEF}$depth$, $v_{sh}$, $v_{sv}$ & \cellcolor[HTML]{EFEFEF}12,782 $\times$ 3 $\times$ 81      \\
                                                                            & \cellcolor[HTML]{FFFFFF}\citet{fichtner_2015_Crust}                                                                             & \cellcolor[HTML]{FFFFFF}CSEM-Iberian           & \cellcolor[HTML]{FFFFFF}\begin{tabular}[c]{@{}c@{}}Western\\ Mediterranean\end{tabular} & \cellcolor[HTML]{FFFFFF}$depth$, $v_{sh}$, $v_{sv}$ & \cellcolor[HTML]{FFFFFF}9,102 $\times$ 3 $\times$ 81       \\
                                                                            & \cellcolor[HTML]{EFEFEF}\citet{colli_2013_Full}                                                                                 & \cellcolor[HTML]{EFEFEF}CSEM-South Atlantic    & \cellcolor[HTML]{EFEFEF}South Atlantic                                                  & \cellcolor[HTML]{EFEFEF}$depth$, $v_{sh}$, $v_{sv}$ & \cellcolor[HTML]{EFEFEF}7,371 $\times$ 3 $\times$ 51       \\
                                                                            & \cellcolor[HTML]{FFFFFF}\begin{tabular}[c]{@{}c@{}}\citet{rickers_2013_Iceland} \\ \citet{krischer_2018_Automated}\end{tabular} & \cellcolor[HTML]{FFFFFF}CSEM-North Atlantic    & \cellcolor[HTML]{FFFFFF}North Atlantic                                                  & \cellcolor[HTML]{FFFFFF}$depth$, $v_{sh}$, $v_{sv}$ & \cellcolor[HTML]{FFFFFF}14,541 $\times$ 3 $\times$ 51      \\
                                                                            & \cellcolor[HTML]{EFEFEF}\citet{simutė_2016_Fullwaveform}                                                                        & \cellcolor[HTML]{EFEFEF}CSEM-Japan             & \cellcolor[HTML]{EFEFEF}Japanese Island                                                 & \cellcolor[HTML]{EFEFEF}$depth$, $v_{sh}$, $v_{sv}$ & \cellcolor[HTML]{EFEFEF}14,641 $\times$ 3 $\times$ 61      \\
\multirow{-14}{*}{\begin{tabular}[c]{@{}c@{}}OpenSWI\\ deep\end{tabular}}   & \cellcolor[HTML]{FFFFFF}\begin{tabular}[c]{@{}c@{}}\citet{fichtner_2009_Full}\\ \citet{fichtner_2010_Full}\end{tabular}         & \cellcolor[HTML]{FFFFFF}CSEM-Astralasia        & \cellcolor[HTML]{FFFFFF}Australasian                                                    & \cellcolor[HTML]{FFFFFF}$depth$, $v_{sh}$, $v_{sv}$ & \cellcolor[HTML]{FFFFFF}4,131 $\times$ 3 $\times$ 51       \\ \hline
                                                                            & \cellcolor[HTML]{EFEFEF}\citet{fu_2022_Improved}                                                                                & \cellcolor[HTML]{EFEFEF}LongBeach              & \cellcolor[HTML]{EFEFEF}USA                                                             & \cellcolor[HTML]{EFEFEF}$depth$, $v_s$              & \cellcolor[HTML]{EFEFEF}5,297 $\times$ 2 $\times$ 241      \\
\multirow{-2}{*}{\begin{tabular}[c]{@{}c@{}}OpenSWI\\ real\end{tabular}}    & \cellcolor[HTML]{FFFFFF}\citet{xiao_2024_CSRM10}                                                                                & \cellcolor[HTML]{FFFFFF}CSRM                   & \cellcolor[HTML]{FFFFFF}Continental China                                               & \cellcolor[HTML]{FFFFFF}$depth$, $v_s$              & \cellcolor[HTML]{FFFFFF}12,901 $\times$ 2 $\times$ 145     \\ \hline
\end{tabular}
\label{table1:data_source}
\end{table}

\subsubsection{Extraction and Parameterization of 1-D Velocity Profiles}
\label{sec:extraction_and_parameterization}

\begin{figure*}[ht]
    \centering
    \includegraphics[width=1.0\textwidth]{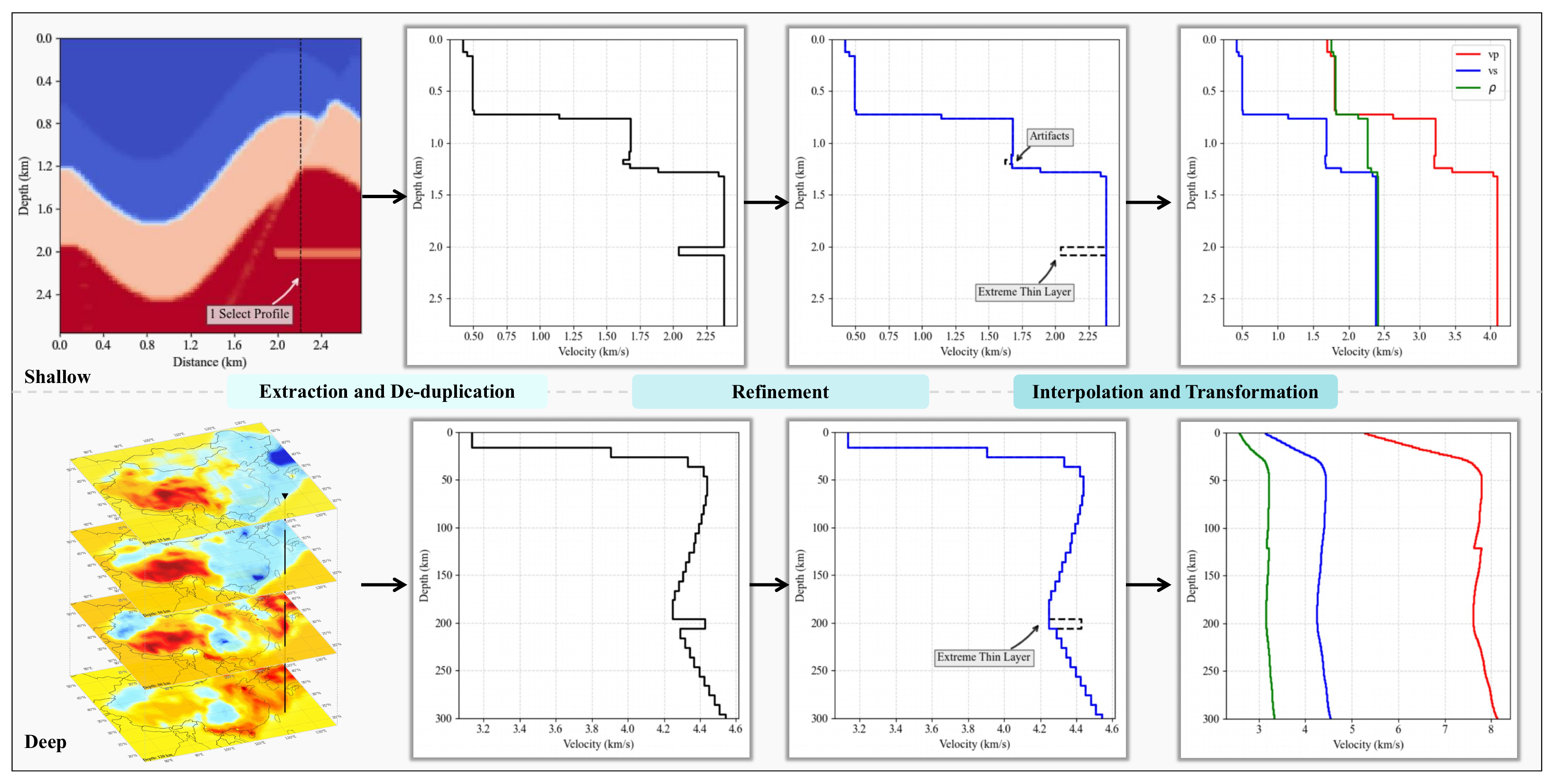}
    \caption{Workflow for extracting and parameterizing 1-D velocity profiles. The upper row shows the process for OpenSWI-shallow, derived from multiple 2-D geological cross-sections, while the lower row illustrates the process for OpenSWI-deep, based on curated 3-D geological models. The workflow includes profile extraction, de-duplication, structure refinement, interpolation, standardization, and parameter conversion to generate depth, \(v_s\) (blue), \(v_p\) (red), and \(\rho\) (green) for forward modeling of surface wave dispersion curves.}
    \label{fig2:1d_profile_extraction}
\end{figure*}

    After completing the quality control and standardization of the geological models, the next step was to construct 1-D velocity profiles suitable for forward modeling. As illustrated in Figure~\ref{fig2:1d_profile_extraction}, this process involved multiple stages, including profile extraction from 2-D or 3-D geological models, removal of redundant samples, structural rationalization, and parameter completion.
    
    Each 1-D profile contains key physical parameters extending from the surface to the target depth range, including depth, $S$-wave velocity ($v_s$), $P$-wave velocity ($v_p$), and density ($\rho$). The procedure is described as follows:
    \begin{enumerate}

        \item \textbf{Extraction and de-duplication of 1-D profiles}: Vertical 1-D \(v_s\) profiles were extracted from 2-D geological cross-sections and 3-D geological models at each surface grid point (or a selected subset of grid points). In some geological models, particularly those with horizontally layered structures, duplicate or nearly identical profiles were identified. These duplicates were removed to ensure each sample was both unique and representative.

        \item \textbf{Structure refinement of 1-D profiles}: To enhance numerical stability during forward modeling, abnormally thin layers or those with extreme velocity values were merged or smoothed. This step minimized non-physical artifacts in the simulations and improved the physical plausibility of the 1-D models.
    
        \item \textbf{Interpolation and standardization}: Uniform layer-thickness interpolation was applied to ensure model consistency across different application scenarios. For shallow subsurface models, layers were resampled at 40~m intervals, whereas for deep-Earth models, a coarser 1~km interval was adopted. This standardization facilitated large-scale batch processing and streamlined integration with deep learning frameworks. We note, however, that some studies may prefer non-uniform layer-thickness schemes (e.g., finer resolution in the shallow part and coarser resolution at greater depths). To support such flexibility, users can easily regenerate alternative dataset versions using the original construction scripts we provide.
    
        \item \textbf{Completion of Other Physical Parameters}: To derive complete physical models, \(v_p\) and \(\rho\) were computed as follows. For depths \(< 120\) km, empirical relationships from \citet{brocher_2005_Empirical} were used to estimate \(v_p\) and \(\rho\) based on known \(v_s\) values, ensuring physical consistency. For depths \(\geq 120\) km, where empirical formulas are less applicable, a constant Poisson's ratio of 1.79 was assumed to calculate \(v_p\) from \(v_s\), with \(\rho\) subsequently derived from \(v_p\) using Brocher's empirical relationship.

    \end{enumerate}
    
    Through these steps, we generated a comprehensive collection of 1-D velocity profiles characterized by geological diversity, physical consistency, and numerical stability.

\subsubsection{Augmentation of Velocity Models for Geological Diversity}
\label{sec:augmentation}

\begin{figure}[!ht]
    \centering
    \includegraphics[width=1.0\textwidth]{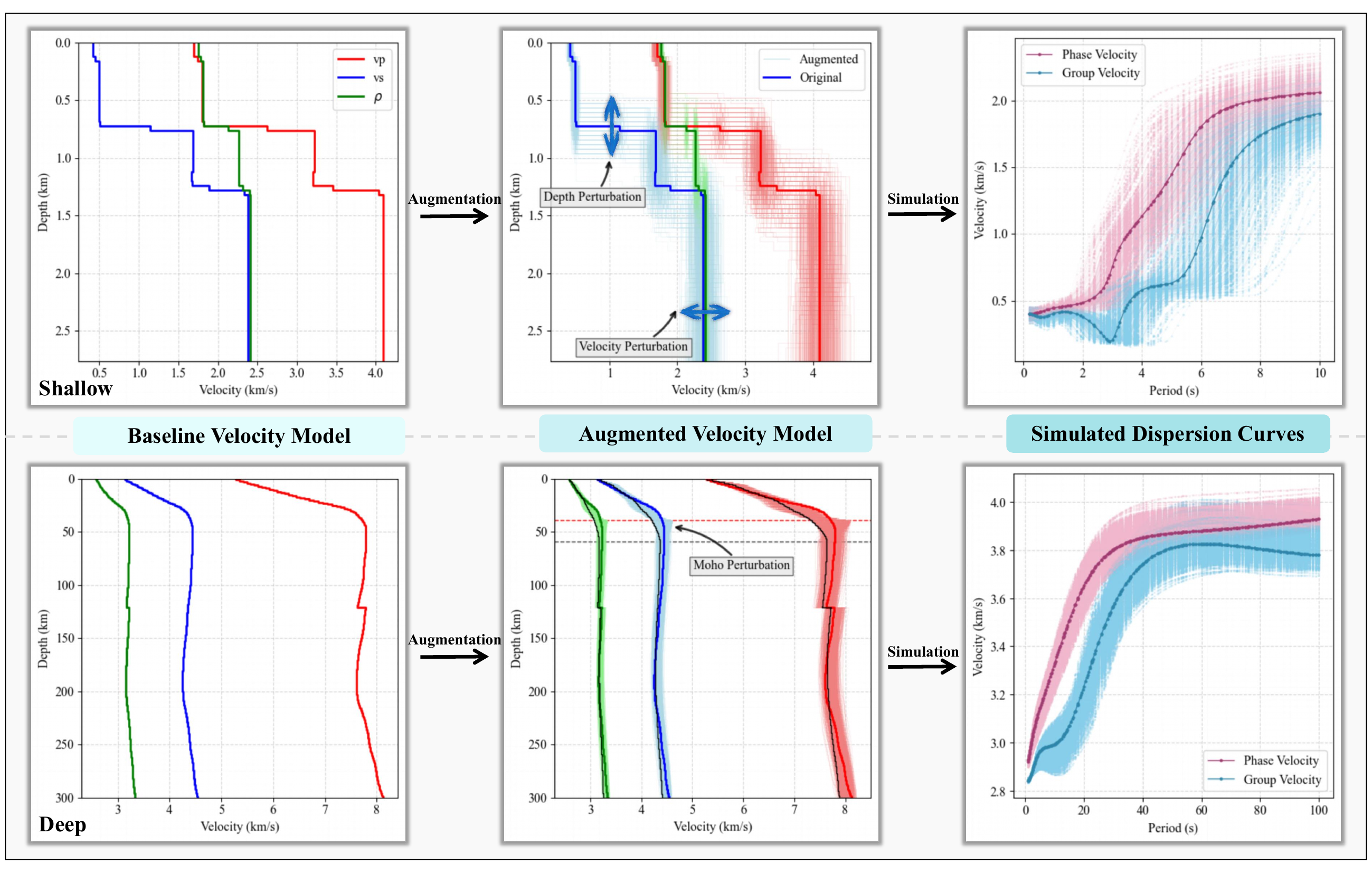}
    \caption{Illustration of data augmentation and forward simulation examples. The top row shows perturbation-based augmentation applied to OpenSWI-shallow data, which increases variability in shallow 1-D velocity profiles. The bottom row shows feature-aware perturbation applied to OpenSWI-deep data, focusing on key structural features such as the Moho discontinuity. Thick lines represent the original 1-D profiles and their corresponding dispersion curves, while thin lines represent the augmented profiles and dispersion curves.}
    \label{fig3:data_augmentation}
\end{figure}

    Although the 1-D velocity profiles extracted and transformed from the aforementioned 2-D and 3-D geological models already surpass those used in previous studies in both quantity and diversity, they still cannot fully capture the complete range of geological types and their characteristic variations. To further broaden the dataset’s representativeness and establish a scalable data construction workflow, we designed and implemented multiple data augmentation strategies based on the original 2-D geological profiles and the processed 1-D velocity models, as outlined below:

    \begin{enumerate}

        \item \textbf{Perturbation-based augmentation of shallow 1-D velocity profiles:} For near-surface geological models, controlled perturbations were applied to both velocities and layer thicknesses while preserving the overall layer structure, thereby enhancing variability across different geological scenarios. The procedure includes: (1) extracting the primary layers from the 1-D profiles; (2) applying random perturbations with constrained amplitudes to the velocity and thickness of each layer; and (3) performing structural plausibility checks on the perturbed profiles, followed by interpolation and parameter conversion as detailed in Section~\ref{sec:extraction_and_parameterization} to ensure physical and numerical validity. The top row of Figure~\ref{fig3:data_augmentation} illustrates the augmentation workflow and the resulting variations for a representative 1-D profile.
    
        \item \textbf{Feature-aware augmentation of deep 1-D velocity profiles:} For deep geological structures characterized by distinct geophysical interfaces (e.g., the Moho discontinuity), we implemented a feature-aware perturbation strategy to improve model sensitivity to key geological boundaries. The procedure involves: (1) identifying the Moho interface in each 1-D profile; (2) fitting the crustal and mantle layers above and below the interface with cubic spline functions, where the number of spline nodes is randomly selected between 3–6 and 8–12, respectively; and (3) applying random perturbations to the velocity values at the spline nodes, followed by curve smoothing and re-interpolation to generate new deep velocity profiles. The bottom row of Figure~\ref{fig3:data_augmentation} illustrates the complete workflow and resulting variations for a representative 1-D profile.
    
        \item \textbf{Generative-model-based augmentation of 2-D geological models:} To further enrich geological feature diversity and enable scalable dataset expansion tailored to user needs, we employed deep generative techniques, such as diffusion probabilistic models (DDPMs, \citet{ho_2020_Denoising}), using the 2-D geological cross-section data collected in Section~\ref{sec:collection_and_QC}. These models learn spatial feature distributions and synthesize additional 2-D geological models with improved geological consistency and structural diversity. This component of the workflow is described in greater detail in the subsequent section on shallow-subsurface dataset construction.

    \end{enumerate}
    
    These augmentation strategies substantially enriched the dataset in terms of geological types, structural complexity, and the representation of key features. As a result, they provide deep learning models with more diverse and comprehensive training samples, thereby improving generalization and robustness when applied to complex geological settings.

\subsubsection{Forward Modeling of Surface-wave Dispersion Curves}
\label{sec:forward_modeling}
    Based on the constructed 1-D velocity profiles, we employed efficient geophysical forward modeling tools to generate the corresponding surface-wave dispersion curves. Forward modeling is a critical step in dataset construction, ensuring that the simulated dispersion curves faithfully capture the propagation characteristics of surface waves in different subsurface media. The workflow comprises three main components:

    \begin{itemize}
        \item \textbf{Defining the period range of dispersion curves:} In practice, the period range and sampling points of observed dispersion curves vary considerably. To enhance the diversity and applicability of the dataset, we designed a hybrid sampling strategy for constructing the period axis. This strategy integrates uniform, random, and logarithmic sampling, with increased sampling density in the high-frequency range \citep{wang_2023_Deeplearningbased,liu_2025_DispFormer}. Such design ensures broad coverage of surface-wave responses across different period bands, improving both the representativeness and utility of the simulated data.

        \item \textbf{Forward computation of dispersion curves:} The forward modeling of surface wave dispersion curves fundamentally involves numerically solving the dispersion equation across a range of frequencies ($f$), where frequency is defined as the reciprocal of the period $T$ (i.e., $f = 1/T$), to determine the corresponding phase velocity $c$ for each mode \citep{thomson_1950_Transmission, haskell_1953_Dispersion, liu_2024_Multimodal}. This process can be formulated as a root-finding problem for the dispersion function $D$:
        \begin{equation}
            D(c, f, \mathbf{m}) = 0,
        \end{equation}
        where $\mathbf{m}$ denotes the elastic parameters of the layered medium, and $D$ encapsulates the frequency-dependent behavior of wave propagation in this structure. Solving Eq. (1) for each frequency yields the phase velocity dispersion curve $c(f)$ (also denoted as $v_{\mathrm{phase}}$), which characterizes the propagation speed of each harmonic component of the wavefield.
        
        In addition to phase velocity, the group velocity $v_{\mathrm{group}}$, which describes the propagation speed of wave packets, is a critical quantity for surface wave analysis. It is obtained as the derivative of angular frequency $\omega = 2\pi f$ with respect to wavenumber $k$, and can be expressed in terms of the phase velocity as:
        \begin{equation}
            v_{\mathrm{group}} = \frac{d \omega}{d k} = c(f) - f \frac{d c(f)}{d f}.
        \end{equation}
        The group velocity curve complements the phase velocity curve by offering additional sensitivity to subsurface structure and is especially useful in tomographic and inversion applications where energy transport characteristics are of interest.
        
        For each 1-D velocity model, we used the Python library \texttt{Disba} (\url{https://keurfonluu.github.io/disba/}), adapted from the classical seismological software package \texttt{Computer Programs in Seismology (CPS)} \citep{herrmann_2013_Computer}, to compute the dispersion curves. This tool efficiently calculates the fundamental-mode phase-velocity and group-velocity characteristics of Rayleigh waves and outputs complete period–velocity pairs (period, phase velocity, and group velocity) for each velocity model, ensuring comprehensive information for inversion tasks.
    
        \item \textbf{Parallelization and computational acceleration:} Given the large scale of the dataset, we implemented multi-process parallelization and matrix-based batch processing to significantly improve computational efficiency. These optimizations enabled the simulation of hundreds of thousands to millions of dispersion curves within a practical timeframe, meeting the data requirements of deep learning applications.
    \end{itemize}
    
    This workflow produced a large-scale, quality-controlled dataset of surface-wave dispersion curves. Figure~\ref{fig3:data_augmentation} showcases examples of dispersion curves from the OpenSWI-shallow and OpenSWI-deep datasets. These simulated data provide a solid foundation for training deep learning–based inversion models, facilitating applications in resource exploration and imaging of Earth’s internal structure.

\begin{table}[ht]
\renewcommand{\arraystretch}{1.25}
\caption{Comprehensive summary of the OpenSWI dataset, describing its categories (OpenSWI-shallow, OpenSWI-deep, and OpenSWI-real), associated period ranges (seconds, s), depth ranges (kilometers, km), and sampling intervals (kilometers, km), as well as the extracted and augmented 1-D velocity profiles ($depth$, $v_p$, $v_s$, $\rho$), expressed as $N$~profiles~$\times$~$M$~model~variables~$\times$~$L$~layers.}

\begin{tabular}{cccccc}
\hline
Group                                                                                            & Datasets                                       & Period Range (s)                      & \begin{tabular}[c]{@{}c@{}}Depth  Range (km)/\\ Depth Interval (km)\end{tabular} & \begin{tabular}[c]{@{}c@{}}Extracted 1-D \\ Velocity Profiles\end{tabular} & \begin{tabular}[c]{@{}c@{}}Augmented 1-D \\ Velocity Profiles\end{tabular} \\ \hline
\multicolumn{1}{c|}{}                                                                            & \cellcolor[HTML]{FFFFFF}Flat                   & \cellcolor[HTML]{FFFFFF}0.1-10        & \cellcolor[HTML]{FFFFFF}0-2.8 / 0.04                                             & \cellcolor[HTML]{FFFFFF}29,379 $\times$ 4 $\times$ 70                                    & \cellcolor[HTML]{FFFFFF}1,490,415 $\times$ 4 $\times$ 70                                 \\
\multicolumn{1}{c|}{}                                                                            & \cellcolor[HTML]{EFEFEF}Flat+Fault             & \cellcolor[HTML]{EFEFEF}0.1-10        & \cellcolor[HTML]{EFEFEF}0-2.8 / 0.04                                             & \cellcolor[HTML]{EFEFEF}292,933 $\times$ 4 $\times$ 70                                   & \cellcolor[HTML]{EFEFEF}2,925,151 $\times$ 4 $\times$ 70                                 \\
\multicolumn{1}{c|}{}                                                                            & \cellcolor[HTML]{FFFFFF}Fold                   & \cellcolor[HTML]{FFFFFF}0.1-10        & \cellcolor[HTML]{FFFFFF}0-2.8 / 0.04                                             & \cellcolor[HTML]{FFFFFF}295,751 $\times$ 4 $\times$ 70                                   & \cellcolor[HTML]{FFFFFF}2,952,975 $\times$ 4 $\times$ 70                                 \\
\multicolumn{1}{c|}{}                                                                            & \cellcolor[HTML]{EFEFEF}Fold+Fault             & \cellcolor[HTML]{EFEFEF}0.1-10        & \cellcolor[HTML]{EFEFEF}0-2.8 / 0.04                                             & \cellcolor[HTML]{EFEFEF}537,751 $\times$ 4 $\times$ 70                                   & \cellcolor[HTML]{EFEFEF}5,369,692 $\times$ 4 $\times$ 70                                 \\
\multicolumn{1}{c|}{}                                                                            & \cellcolor[HTML]{FFFFFF}Field                  & \cellcolor[HTML]{FFFFFF}0.1-10        & \cellcolor[HTML]{FFFFFF}0-2.8 / 0.04                                             & \cellcolor[HTML]{FFFFFF}2,338,248 $\times$ 4 $\times$ 70                                 & \cellcolor[HTML]{FFFFFF}9,345,103 $\times$ 4 $\times$ 70                                 \\
\multicolumn{1}{c|}{\multirow{-6}{*}{\begin{tabular}[c]{@{}c@{}}OpenSWI\\ shallow\end{tabular}}} & \cellcolor[HTML]{EFEFEF}All                    & \cellcolor[HTML]{EFEFEF}0.1-10        & \cellcolor[HTML]{EFEFEF}0-2.8 / 0.04                                             & \cellcolor[HTML]{EFEFEF}3,494,062 $\times$ 4 $\times$ 70                                 & \cellcolor[HTML]{EFEFEF}22,083,336x 4 $\times$ 70                                 \\ \hline
\multicolumn{1}{c|}{}                                                                            & \cellcolor[HTML]{FFFFFF}LITHO1.0               & \cellcolor[HTML]{FFFFFF}1-100         & \cellcolor[HTML]{FFFFFF}0-300 / 1.0                                              & \cellcolor[HTML]{FFFFFF}40,959 $\times$ 4 $\times$ 300                                   & \cellcolor[HTML]{FFFFFF}24,5771 $\times$ 4 $\times$ 70                                   \\
\multicolumn{1}{c|}{}                                                                            & \cellcolor[HTML]{EFEFEF}USTClitho1.0           & \cellcolor[HTML]{EFEFEF}1-100         & \cellcolor[HTML]{EFEFEF}0-300 / 1.0                                              & \cellcolor[HTML]{EFEFEF}9,125 $\times$ 4 $\times$ 300                                    & \cellcolor[HTML]{EFEFEF}54,750 $\times$ 4 $\times$ 70                                    \\
\multicolumn{1}{c|}{}                                                                            & \cellcolor[HTML]{FFFFFF}Central-and-Western US & \cellcolor[HTML]{FFFFFF}1-100         & \cellcolor[HTML]{FFFFFF}0-300 / 1.0                                              & \cellcolor[HTML]{FFFFFF}6,803 $\times$ 4 $\times$ 300                                    & \cellcolor[HTML]{FFFFFF}40,818 $\times$ 4 $\times$ 70                                    \\
\multicolumn{1}{c|}{}                                                                            & \cellcolor[HTML]{EFEFEF}Continental China      & \cellcolor[HTML]{EFEFEF}1-100         & \cellcolor[HTML]{EFEFEF}0-300 / 1.0                                              & \cellcolor[HTML]{EFEFEF}4,516 $\times$ 4 $\times$ 300                                    & \cellcolor[HTML]{EFEFEF}27,096 $\times$ 4 $\times$ 70                                    \\
\multicolumn{1}{c|}{}                                                                            & \cellcolor[HTML]{FFFFFF}US Upper-Mantle        & \cellcolor[HTML]{FFFFFF}1-100         & \cellcolor[HTML]{FFFFFF}0-300 / 1.0                                              & \cellcolor[HTML]{FFFFFF}3,678 $\times$ 4 $\times$ 300                                    & \cellcolor[HTML]{FFFFFF}22,061 $\times$ 4 $\times$ 70                                    \\
\multicolumn{1}{c|}{}                                                                            & \cellcolor[HTML]{EFEFEF}EUcrust                & \cellcolor[HTML]{EFEFEF}1-100         & \cellcolor[HTML]{EFEFEF}0-300 / 1.0                                              & \cellcolor[HTML]{EFEFEF}43,520 $\times$ 4 $\times$ 300                                   & \cellcolor[HTML]{EFEFEF}261,155 $\times$ 4 $\times$ 70                                   \\
\multicolumn{1}{c|}{}                                                                            & \cellcolor[HTML]{FFFFFF}Alaska                 & \cellcolor[HTML]{FFFFFF}1-100         & \cellcolor[HTML]{FFFFFF}0-300 / 1.0                                              & \cellcolor[HTML]{FFFFFF}19,408 $\times$ 4 $\times$ 300                                   & \cellcolor[HTML]{FFFFFF}116,448 $\times$ 4 $\times$ 70                                   \\
\multicolumn{1}{c|}{}                                                                            & \cellcolor[HTML]{EFEFEF}CSEM-Europe            & \cellcolor[HTML]{EFEFEF}1-100         & \cellcolor[HTML]{EFEFEF}0-300 / 1.0                                              & \cellcolor[HTML]{EFEFEF}21,931 $\times$ 4 $\times$ 300                                   & \cellcolor[HTML]{EFEFEF}131,586 $\times$ 4 $\times$ 70                                   \\
\multicolumn{1}{c|}{}                                                                            & \cellcolor[HTML]{FFFFFF}CSEM-Eastmed           & \cellcolor[HTML]{FFFFFF}1-100         & \cellcolor[HTML]{FFFFFF}0-300 / 1.0                                              & \cellcolor[HTML]{FFFFFF}12,782 $\times$ 4 $\times$ 300                                   & \cellcolor[HTML]{FFFFFF}76,692 $\times$ 4 $\times$ 70                                    \\
\multicolumn{1}{c|}{}                                                                            & \cellcolor[HTML]{EFEFEF}CSEM-Iberian           & \cellcolor[HTML]{EFEFEF}1-100         & \cellcolor[HTML]{EFEFEF}0-300 / 1.0                                              & \cellcolor[HTML]{EFEFEF}9,102 $\times$ 4 $\times$ 300                                    & \cellcolor[HTML]{EFEFEF}54,612 $\times$ 4 $\times$ 70                                    \\
\multicolumn{1}{c|}{}                                                                            & \cellcolor[HTML]{FFFFFF}CSEM-South Atlantic    & \cellcolor[HTML]{FFFFFF}1-100         & \cellcolor[HTML]{FFFFFF}0-300 / 1.0                                              & \cellcolor[HTML]{FFFFFF}7,371 $\times$ 4 $\times$ 300                                    & \cellcolor[HTML]{FFFFFF}44,226 $\times$ 4 $\times$ 70                                    \\
\multicolumn{1}{c|}{}                                                                            & \cellcolor[HTML]{EFEFEF}CSEM-North Atlantic    & \cellcolor[HTML]{EFEFEF}1-100         & \cellcolor[HTML]{EFEFEF}0-300 / 1.0                                              & \cellcolor[HTML]{EFEFEF}14,541 $\times$ 4 $\times$ 300                                   & \cellcolor[HTML]{EFEFEF}87,246 $\times$ 4 $\times$ 70                                    \\
\multicolumn{1}{c|}{}                                                                            & \cellcolor[HTML]{FFFFFF}CSEM-Japan             & \cellcolor[HTML]{FFFFFF}1-100         & \cellcolor[HTML]{FFFFFF}0-300 / 1.0                                              & \cellcolor[HTML]{FFFFFF}14,641 $\times$ 4 $\times$ 300                                   & \cellcolor[HTML]{FFFFFF}87,846 $\times$ 4 $\times$ 70                                    \\
\multicolumn{1}{c|}{}                                                                            & \cellcolor[HTML]{EFEFEF}CSEM-Astralasia        & \cellcolor[HTML]{EFEFEF}1-100         & \cellcolor[HTML]{EFEFEF}0-300 / 1.0                                              & \cellcolor[HTML]{EFEFEF}4,131 $\times$ 4 $\times$ 300                                    & \cellcolor[HTML]{EFEFEF}24,786 $\times$ 4 $\times$ 70                                    \\
\multicolumn{1}{c|}{\multirow{-15}{*}{\begin{tabular}[c]{@{}c@{}}OpenSWI\\ deep\end{tabular}}}   & \cellcolor[HTML]{FFFFFF}All                    & \cellcolor[HTML]{FFFFFF}1-100         & \cellcolor[HTML]{FFFFFF}0-300 / 1.0                                              & \cellcolor[HTML]{FFFFFF}212,508 $\times$ 4 $\times$ 300                                  & \cellcolor[HTML]{FFFFFF}1,275,093 $\times$ 4 $\times$ 70                                 \\ \hline
\multicolumn{1}{c|}{}                                                                            & \cellcolor[HTML]{EFEFEF}LongBeach              & \cellcolor[HTML]{EFEFEF}0.263 - 1.666 & \cellcolor[HTML]{EFEFEF}0-1.4  / 0.04                                            & \cellcolor[HTML]{EFEFEF}5,297 $\times$ 4 $\times$ 35                                     & \cellcolor[HTML]{EFEFEF}-                                                  \\
\multicolumn{1}{c|}{\multirow{-2}{*}{\begin{tabular}[c]{@{}c@{}}OpenSWI\\ real\end{tabular}}}    & \cellcolor[HTML]{FFFFFF}CSRM                   & \cellcolor[HTML]{FFFFFF}8 - 70        & \cellcolor[HTML]{FFFFFF}0-120  / 1.0                                             & \cellcolor[HTML]{FFFFFF}12,901 $\times$ 4 $\times$ 120                                   & \cellcolor[HTML]{FFFFFF}-                                                  \\ \hline
\end{tabular}

\label{table2:openswi_details}
\end{table}

\subsubsection{Open-source Implementation}
\label{sec:swidp_opensource}
    To promote reproducibility, scalability, and community engagement, we developed a standardized Python toolkit named \texttt{SWIDP} (Surface Wave Inversion Dataset Preparation pipeline). Built upon the key procedures described in Sections~\ref{sec:collection_and_QC}–\ref{sec:forward_modeling}, \texttt{SWIDP} encapsulates core functionalities such as the extraction and parameterization of 1-D velocity profiles, data augmentation, and large-scale dispersion curve simulation. 
    
    By automating these processes, it enhances the efficiency, transparency, and consistency of dataset preparation. Designed with a modular architecture, \texttt{SWIDP} allows users to flexibly reuse or extend specific components, facilitating seamless adaptation to diverse research needs. Example codes are provided in Appendix~\ref{sec:appendix_swidp_shallow} and \ref{sec:appendix_swidp_deep}. The full source code is openly available at \url{https://github.com/liufeng2317/OpenSWI}, enabling both academic and industrial users to adopt and further develop the toolkit.

%
%

\subsection{OpenSWI-shallow: Large-scale Benchmark for Complex Shallow Geology}
\label{sec:openswi_shallow}

\subsubsection{Building Geological Model Foundations from OpenFWI}

    To establish a representative benchmark dataset for shallow-subsurface surface-wave dispersion curve inversion, we constructed a comprehensive collection of 2-D velocity models with diverse geological structures derived from the OpenFWI dataset \citep{deng_2021_OpenFWI}. These models encompass five primary geological categories: flat layers (Flat), flat layers with faults (Flat–Fault), folded layers (Fold), folded layers with faults (Fold–Fault), and field-style models (Field) inspired by realistic field observations. Each category contains approximately 30,000, 54,000, 30,000, 54,000, and 67,000 models, respectively. All models share a grid resolution of $70 \times 70$ with a spatial sampling interval of 40~m, ensuring sufficient detail to capture the complexity and variability of shallow-subsurface geological features and thereby providing a robust foundation for building a high-quality dataset.

    Based on these 2-D models, we systematically extracted a large number of 1-D velocity profiles according to the geological characteristics of each geological categories. To enhance the dataset’s diversity and coverage, each original 1-D profile was augmented 4 to 10 times by independently applying perturbations of up to 10\% in layer thickness and 5\% in velocity. Following these perturbations, plausibility checks and interpolation adjustments were performed to ensure physical consistency and numerical stability. The final dataset comprises over 22 million 1-D velocity models spanning all geological categories. Detailed statistics of both the extracted and augmented profile counts for each category are summarized in Table~\ref{table2:openswi_details}.

    Forward modeling of fundamental-mode Rayleigh-wave dispersion curves was then conducted for all 1-D models. Given that the maximum model depth is approximately 2.8~km, the simulated period range was defined from 0.2~s to 10~s, with 100 period points sampled per curve. To improve period coverage and model generalization capability, the sampling points were selected using a hybrid strategy combining uniform, random, and logarithmic sampling, contributing 50, 30, and 20 points, respectively. Each dispersion curve includes period, phase-velocity, and group-velocity information, serving as training and validation data for subsequent deep learning applications.

\begin{figure}[ht]
        \centering
        \includegraphics[width=0.65\textwidth]{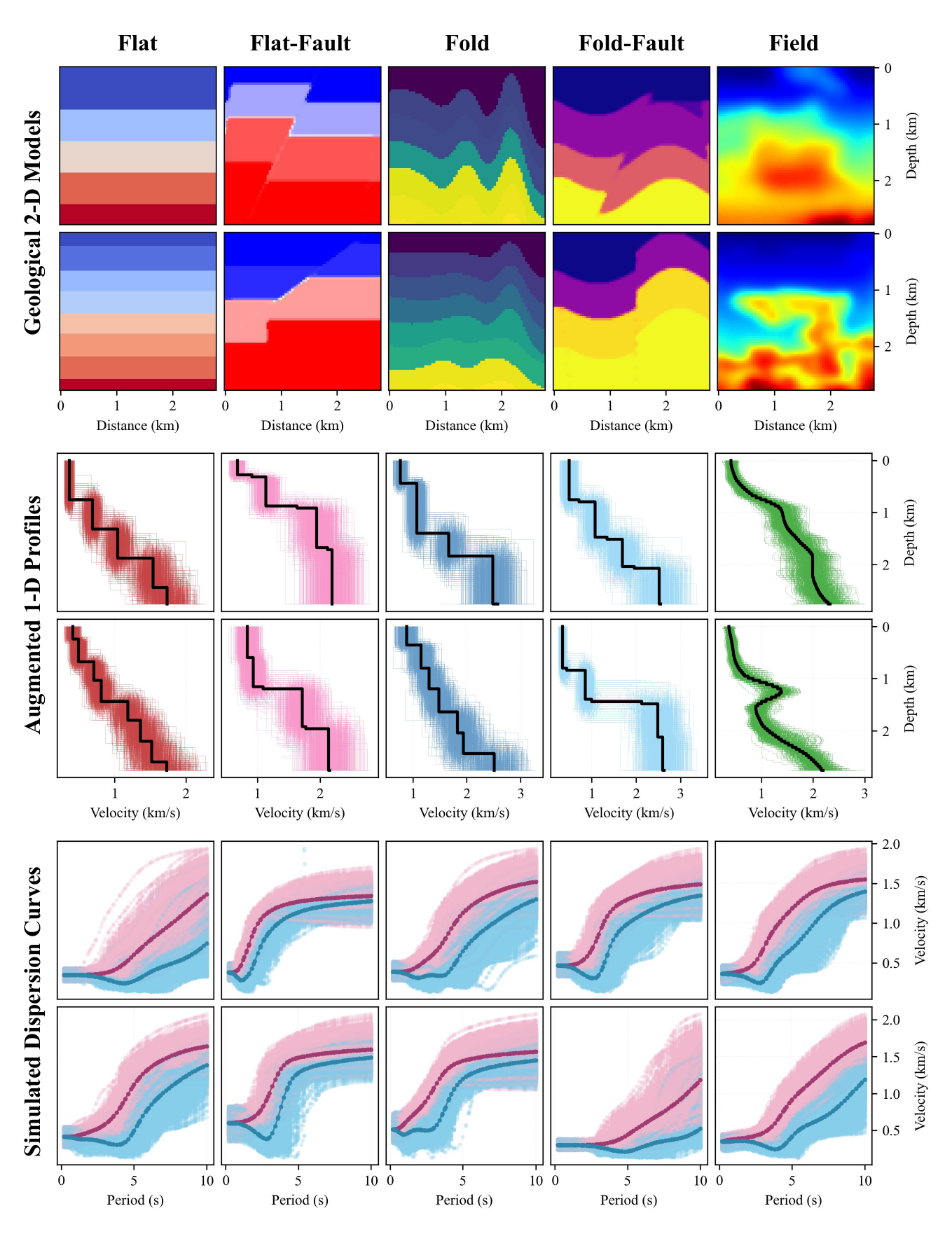}
        \caption{Representative samples from the OpenSWI-shallow dataset. The top two rows present original 2-D velocity models for five geological types: Flat, Flat–Fault, Fold, Fold–Fault, and Field. The middle two rows show the corresponding extracted 1-D velocity profiles (bold black lines) and their augmented variants (thin colored lines). The bottom two rows display the simulated Rayleigh-wave dispersion curves, with phase velocities shown in pink and group velocities in blue.}
        \label{fig4:openswi_shallow_sample}
    \end{figure}

\begin{figure}[ht]
        \centering
        \includegraphics[width=0.8\textwidth]{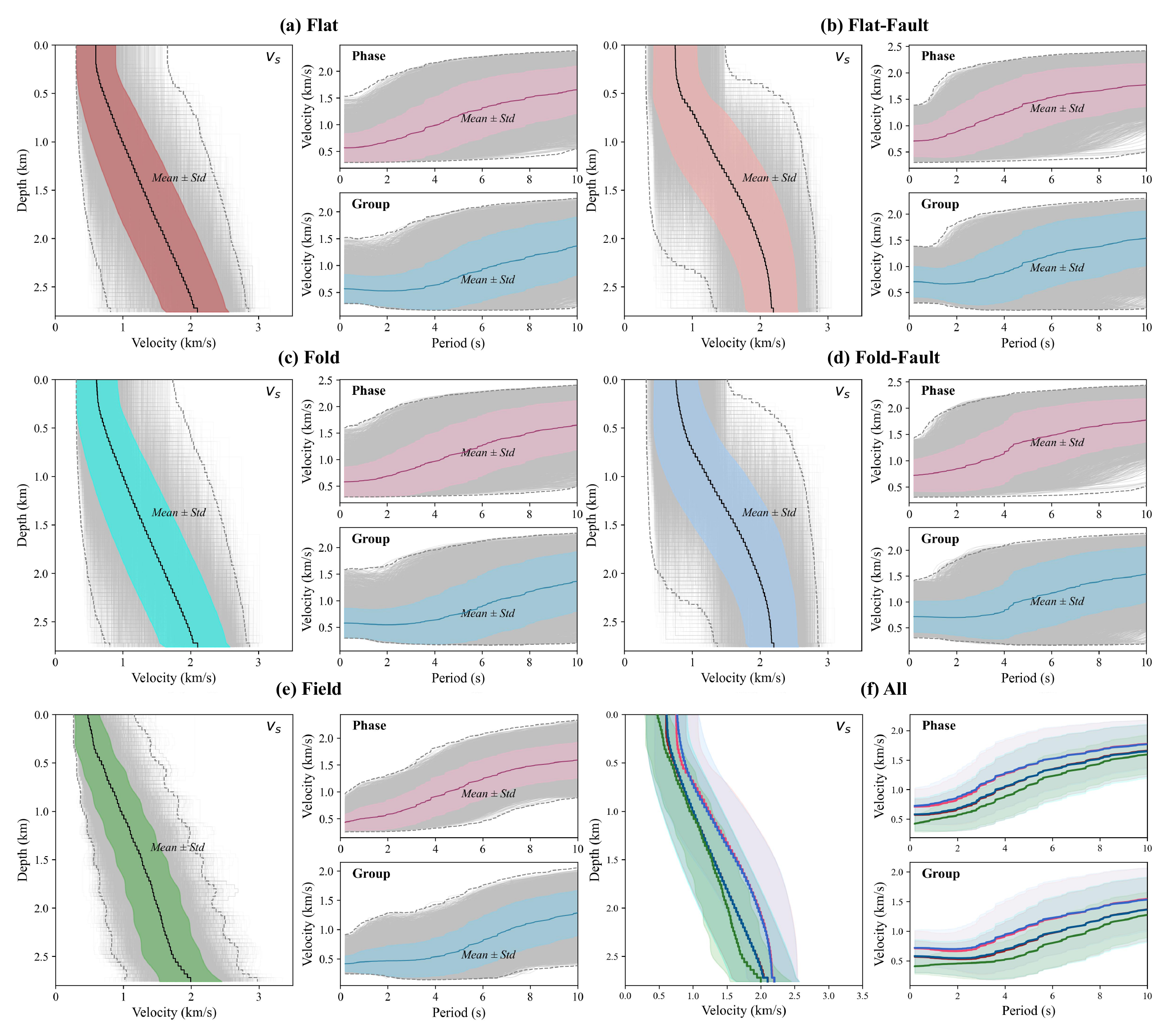}
        \caption{Statistical characteristics of the OpenSWI-shallow dataset. Distribution of 1-D velocity profiles and corresponding dispersion curves for each geological style: (a) Flat, (b) Flat-Fault, (c) Fold, (d) Fold-Fault, (e) Field. The black lines represent the mean, and the shaded regions indicate the $\pm1$ standard deviation range. Panel (f) summarizes the mean and variance across the five geological subsets.}
        \label{fig5:openswi_shallow_statistic}
    \end{figure}

    Figures~\ref{fig4:openswi_shallow_sample} and \ref{fig5:openswi_shallow_statistic} showcase the representativeness and statistical properties of the OpenSWI-shallow dataset. Figure~\ref{fig4:openswi_shallow_sample} illustrates the diverse geological scenarios covered by the dataset through representative 2-D velocity models, systematically extracted 1-D profiles, and their augmented variants, together with the corresponding phase and group velocity dispersion curves. Figure~\ref{fig5:openswi_shallow_statistic} further summarizes the large-scale statistical distributions of profiles and dispersion characteristics across all geological types, highlighting the dataset’s substantial improvements in structural diversity, distributional coverage, and suitability for data-driven surface wave inversion studies.

\clearpage

\subsubsection{Continual Expansion of Geological Structures with DDPM}

    Although the proposed OpenSWI-shallow dataset constructed from OpenFWI substantially improves geological structural diversity compared with existing dispersion curve datasets, it still cannot fully capture the complete range of velocity structure types observed in real subsurface settings. To further enhance the dataset’s scalability in terms of structural complexity and geological representativeness, we incorporated a deep generative module based on DDPMs, specifically designed for the shallow subsurface within the 0–3~km depth range.

    This module uses 2-D velocity models from OpenFWI as training data to develop multiple DDPMs, which learn the distributional characteristics of different geological structures. Starting from Gaussian noise, the DDPMs iteratively generate velocity models with realistic structural features, consistently reproducing faults, folds, and complex sedimentary units. Compared with traditional manual or perturbed augmentation, the DDPM-generated data provide clear advantages in structural continuity, geological realism, and controllable scalability, significantly expanding the foundational velocity model library \citep{ho_2020_Denoising, wang_2023_Prior, taufik_2024_Learned}. Details of the DDPM design and training are provided in Appendix~\ref{sec:appendix_ddpm}, and the code has been publicly released with the SWIDP pipeline for reproducibility.

\begin{figure}[!ht]
        \centering
        \includegraphics[width=0.85\textwidth]{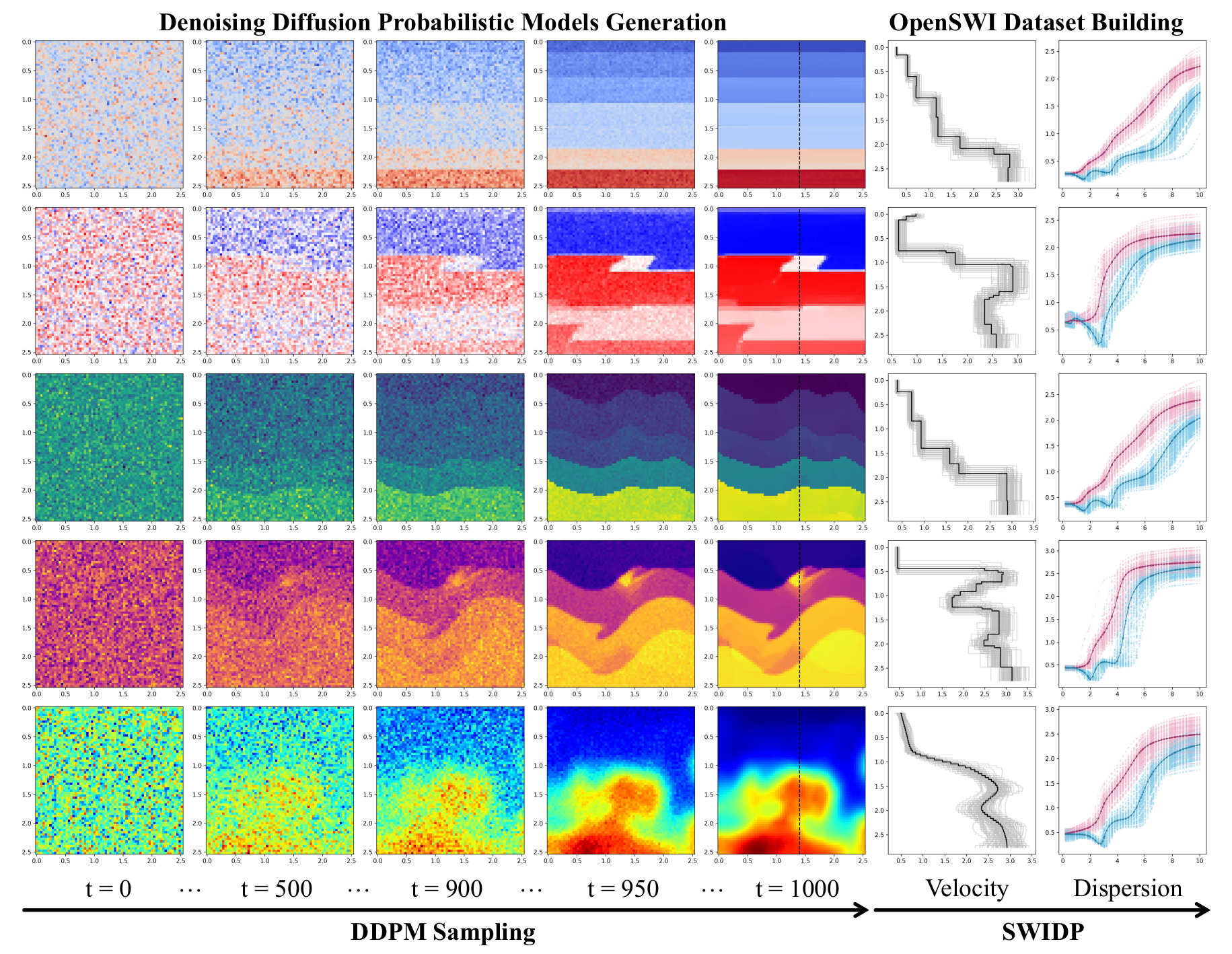}
        \caption{Continual expansion of the OpenSWI-shallow dataset using a diffusion-based generative module. The left panel illustrates a 1000-step denoising trajectory, where Gaussian noise is progressively transformed into 2-D velocity models with realistic geological structures. The right panel presents representative 1-D velocity profiles extracted from the generated models, along with their corresponding Rayleigh-wave dispersion curves simulated using the SWIDP pipeline.}
        \label{fig6:openswi_ddpm}
    \end{figure}

    Figure~\ref{fig6:openswi_ddpm} illustrates the continual expansion of the OpenSWI-shallow dataset using the DDPM module. The diffusion model progressively transforms Gaussian noise into geologically realistic 2-D velocity models through a 1000-step denoising process, from which representative 1-D profiles are extracted and used to simulate Rayleigh-wave dispersion curves. This diffusion-based augmentation strategy substantially enriches the structural diversity and spatial coverage of the dataset, thereby improving the generalization capability of deep learning models.

%
%
\subsection{OpenSWI-deep: Global Coverage Benchmark for Deep Earth Imaging}
\label{sec:openswi_deep}
\begin{figure}[!ht]
        \centering
        \includegraphics[width=1.0\textwidth]{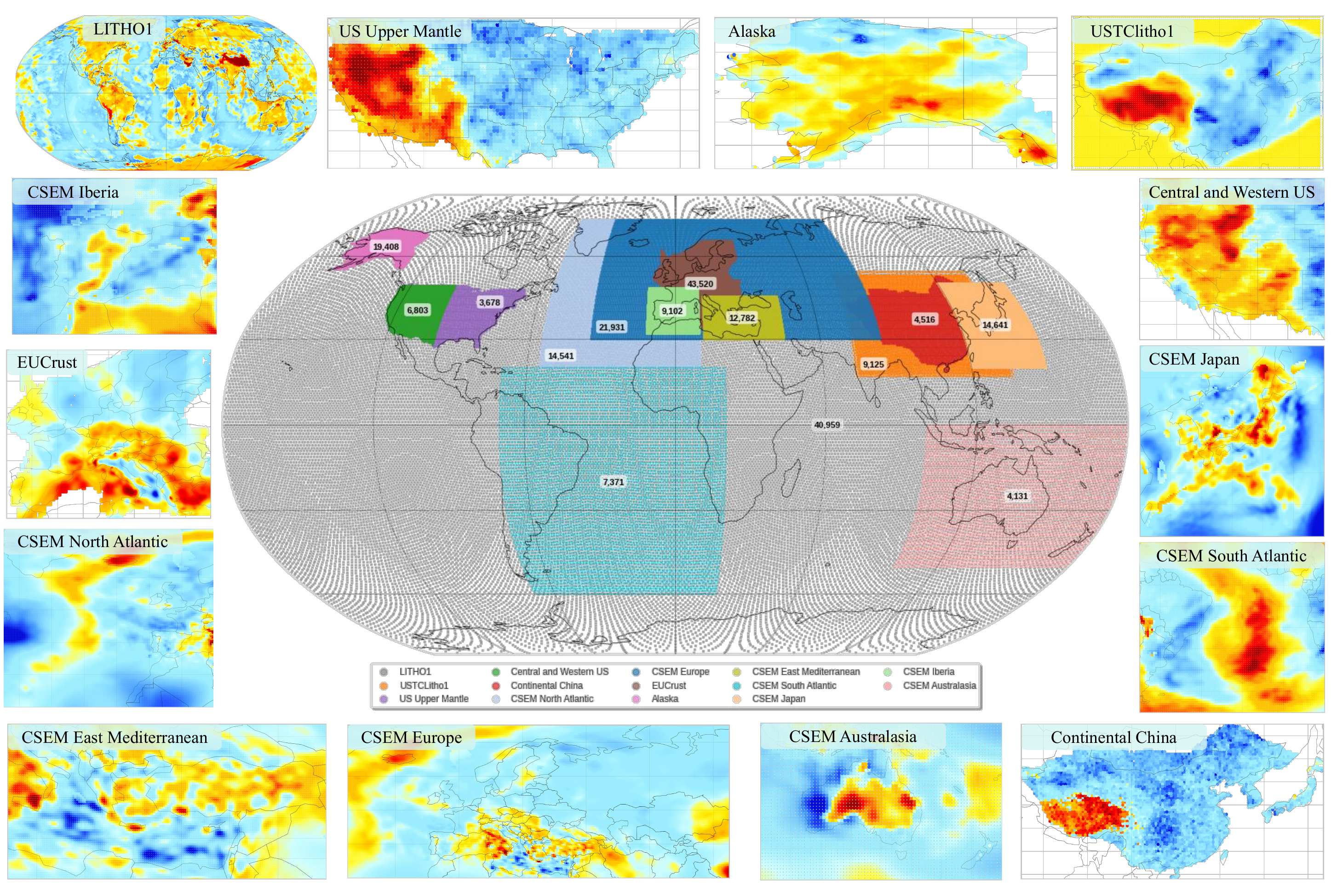}
        \caption{Spatial distribution of the 14 velocity models compiled for the OpenSWI-deep dataset. The collection includes one global-scale model and 13 high-resolution regional models obtained from published literature and geophysical studies. Horizontal slices at a depth of 60~km are shown to illustrate their geographic coverage and tectonic diversity.}
        \label{fig7:openswi_deep_distribution}
    \end{figure}

    Building upon the shallow-subsurface benchmark dataset introduced in Section~\ref{sec:openswi_shallow}, we further extended the OpenSWI framework to deeper Earth structures. However, for the deeper Earth structure, systematic datasets of regular velocity models remain largely unavailable. To address this gap, we compiled a collection of representative 3-D velocity models from published literature and geophysical studies. This collection includes one global-scale model and 13 high-resolution regional models, each constructed using different methodologies and data sources to maximize geological representativeness and geophysical applicability. Figure~\ref{fig7:openswi_deep_distribution} shows the spatial distribution of these 14 models with horizontal slices at a depth of 60~km.

    Among them, LITHO1.0 provides global information on the crust and upper mantle, encompassing sedimentary layers, crust, lithosphere, and asthenosphere, at a spatial resolution of 1\degree\citep{pasyanos_2014_LITHO10}. This model is widely used in seismic tomography and as a reference Earth model. USTClitho1.0, derived from double-difference tomography using seismic data from the Chinese National Seismic Network, resolves crustal and upper mantle structures down to 150~km depth at a horizontal resolution of 0.5\degree, supporting studies of regional deep structures \citep{xin_2019_Highresolution}. The Central and Western US \citep{shen_2013_3D} and Continental China \citep{shen_2016_Seismic} models integrate ambient noise and teleseismic surface waves with receiver function data and apply a Bayesian Monte Carlo inversion to image crust and upper mantle structures to 150~km depth at 0.5\degree resolution. The US Upper Mantle model uses long-period Rayleigh wave ambient noise and Markov chain Monte Carlo inversion to map shear-wave velocities down to 300~km across the continental United States \citep{xie_2018_3D}. Similarly, the EUCrust model, based on four years of ambient noise data from 1,293 broadband stations, resolves the European crust and uppermost mantle with high resolution using Bayesian nonlinear methods \citep{lu_2018_Highresolution}. The Alaska model combines data from over 200 Transportable Array stations and integrates Rayleigh wave ellipticity, phase velocity, and receiver functions, using Markov chain inversion to image structures from the upper mantle to near-surface depths (140~km) \citep{berg_2020_Shear}.
    
    We also included several regional models from The Collaborative Seismic Earth Model Project (CSEM), constructed through full-waveform inversion \citep{fichtner_2006_Adjoint, fichtner_2013_Multiscale, fichtner_2018_Collaborative}. These cover Europe \citep{sabuncu_2017_3D, blom_2020_Seismic}, the Eastern and Western Mediterranean \citep{blom_2020_Seismic, fichtner_2015_Crust}, the South and North Atlantic \citep{colli_2013_Full, rickers_2013_Iceland, krischer_2018_Automated}, the Japanese Islands \citep{simutė_2016_Fullwaveform}, and Australasia \citep{fichtner_2009_Full, fichtner_2010_Full}. These models are characterized by high resolution and structural consistency and are widely used for deep Earth imaging and geodynamic research.
    
    All collected 3-D velocity models underwent quality control, including duplicate removal, anomaly detection, gap interpolation, and format homogenization to ensure consistency for dataset construction. From the processed models, we extracted approximately 210,000 1-D velocity profiles, with detailed statistics for each data source provided in Table~\ref{table2:openswi_details}. To further increase diversity, each profile was augmented five times using a hierarchical strategy: the depth of the Moho discontinuity was first identified, and profiles were divided into crust and upper mantle sections. Each section was parameterized using cubic spline curves, with 3–6 control nodes for the crust and 6–12 nodes for the upper mantle, followed by random perturbations of the nodes to introduce structural variations. This augmentation preserved key geological features (e.g., the Moho interface) while significantly expanding the coverage and variability of the model library, ultimately yielding approximately 1.26 million augmented 1-D velocity models.
    
\begin{figure}[!ht]
        \centering
        \includegraphics[width=1.0\textwidth]{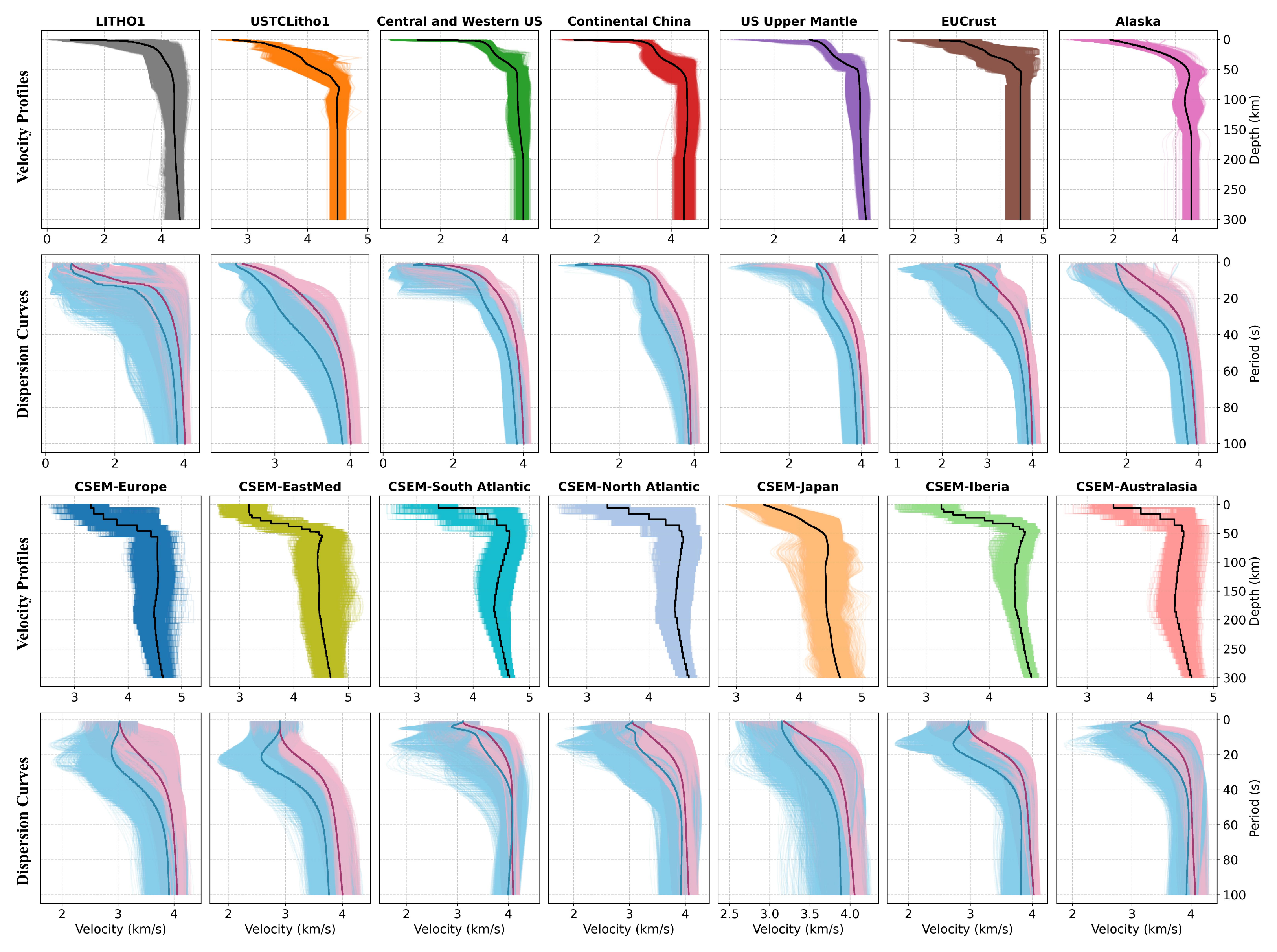}
        \caption{Representative samples from the OpenSWI-deep dataset. The first and third rows show 1-D velocity profiles extracted from the 14 sub-datasets, where the mean velocity model is indicated by a solid black line. The second and fourth rows display the corresponding fundamental-mode Rayleigh-wave dispersion curves over a 1–100~s period range, with the mean phase and group velocities shown in pink and blue, respectively.}
        \label{fig8:openswi_deep_examples}
    \end{figure}

    For each 1-D profile, we simulated fundamental-mode Rayleigh-wave dispersion curves over a 1–100~s period range, sampling 300 points using a combination of uniform, random, and logarithmic strategies (50, 30, and 20 points, respectively). Each dispersion curve, together with its associated velocity profile, constitutes a complete input–output pair for subsequent deep learning model training and validation. Figure~\ref{fig8:openswi_deep_examples} illustrates representative 1-D velocity profiles and their corresponding dispersion curves from the regional models, highlighting the relationships between velocity structures and surface-wave propagation under diverse geological conditions. These results provide high-quality initial data support for global geophysical imaging across different regions and scales.

%
%
\begin{figure}[!ht]
    \centering
    \includegraphics[width=0.85\textwidth]{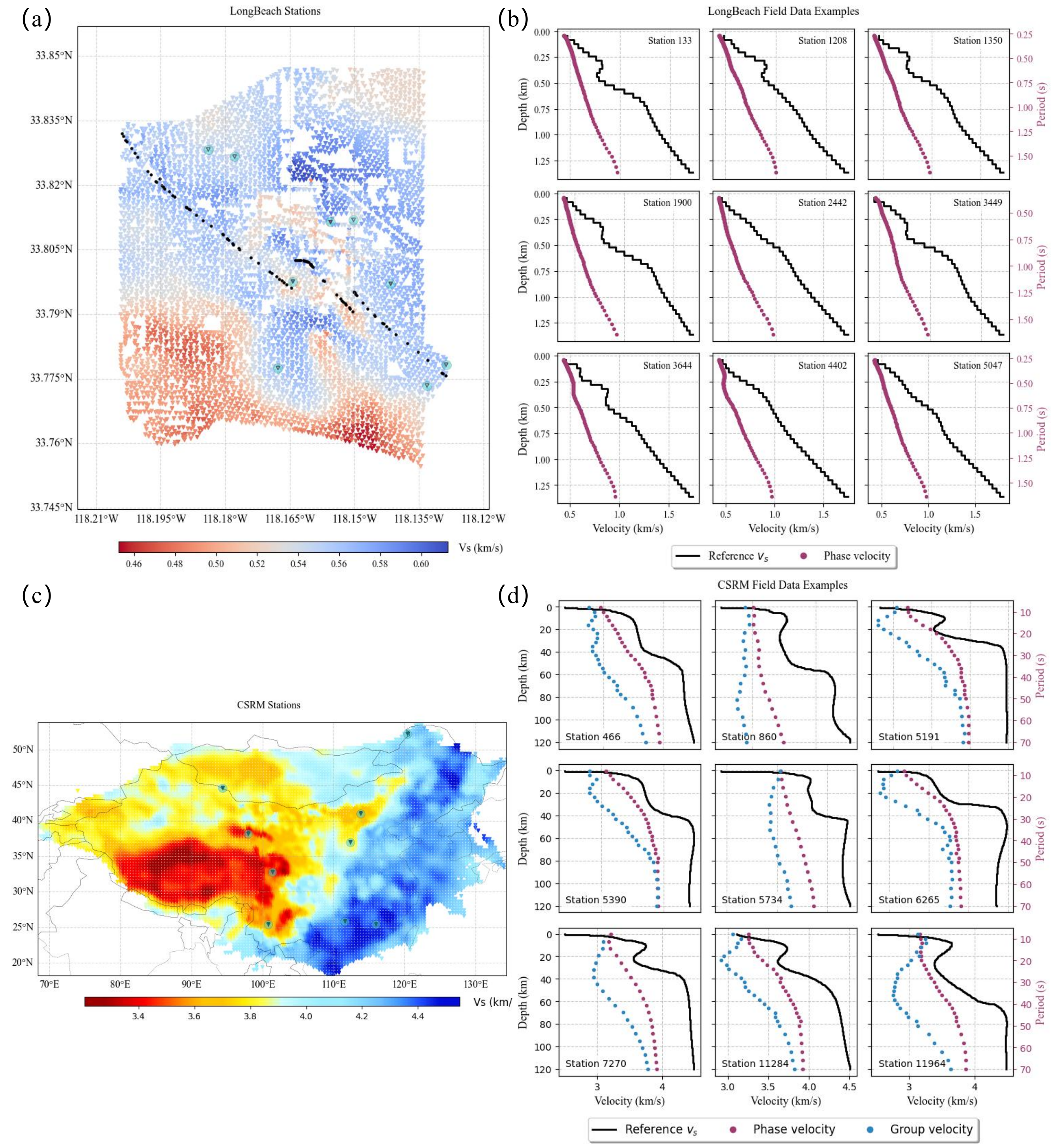}
    \caption{Overview of the OpenSWI-real dataset. (a) Station deployment for the Long Beach dataset in Southern California. (b) Representative observed phase velocity dispersion curves (purple dashed lines) and reference velocity models (black lines) from traditional inversion. (c) Distribution of selected grid points in the CSRM dataset across continental China, with background color denoting velocity at 70~km depth. (d) Representative examples from the CSRM dataset showing observed group (blue) and phase (purple) velocity curves and corresponding reference 1-D velocity profiles (black).}
    \label{fig9:openswi_real_example}
\end{figure}

\subsection{OpenSWI-real: AI-ready Real-world Dataset for Generalization Testing}
\label{sec:openswi_real}

    In addition to the large-scale synthetic velocity profile–dispersion curve datasets designed for model training, we curated multiple AI-ready real-world dispersion curve datasets to assess the adaptability and generalization capability of deep learning methods under practical geophysical conditions.

    The first dataset is derived from the dispersion curve data processed by \citet{fu_2022_Improved} in the Long Beach region of the United States. As shown in Figure~\ref{fig9:openswi_real_example}(a), over 5,200 short-period nodal stations were deployed between January and June 2011, primarily for oilfield surveys \citep{lin_2013_Highresolution}, with an average station spacing of approximately 0.1~km. To achieve adequate spatial resolution, the dense array was divided into multiple subarrays, each with a 2~km radius. Dispersion curves were extracted automatically using a deep neural network after the frequency–Bessel (F–J) transform was applied to compute the frequency–phase velocity spectrum for each subarray. Figure~\ref{fig9:openswi_real_example}(b) shows representative observed dispersion curves from 9 stations (purple dashed lines), together with 1-D reference shear-wave velocity profiles (black solid lines) obtained via traditional inversion methods. This dataset contains only phase velocity data, without group velocity information. After standardized processing, it comprises observed dispersion data from 5,297 stations (period range: 0.263–1.666~s) and corresponding reference velocity models (depth range: 0–1.4km, interpolated at 40~m intervals).

    The second dataset originates from the China Seismological Reference Model Project \citep{wen_2023_China, xiao_2024_CSRM10}. \citet{xiao_2024_CSRM10} collected continuous seismic records from multiple networks, including the China National Seismic Network (CNSN), the China Seismic Array (ChinArray), and the Public Data Management Center (PDMC), spanning 4,196 seismic stations in total. Ambient noise cross-correlations between station pairs produced 639,171 empirical Green’s functions, from which dispersion curves were extracted using frequency–time analysis. Additionally, 54,792 event–station dispersion curves were retrieved from 226 regional seismic events recorded by 1,463 stations. After gridding and quality control, the data were consolidated into 20,514 grid points and standardized to a period range of 8–70~s. To ensure reliability, we retained 12,901 grid points with at least 20 sampled period points. The resulting AI-ready dataset contains observed dispersion curves at these grid points (period range: 8–70~s) and their corresponding reference velocity models (depth range: 0–120~km, interpolated at 1~km intervals). Figure~\ref{fig9:openswi_real_example}(c) shows the spatial distribution of the selected grid points across continental China, with background colors indicating the reference model velocity at 70~km depth. Figure~\ref{fig9:openswi_real_example}(d) presents nine representative grid points, displaying observed dispersion curves (blue: group velocity; purple: phase velocity) and corresponding reference velocity profiles (black solid lines) derived from the traditional inversion results of \citet{xiao_2024_CSRM10}.

%
%
\section{Deep-learning-based Framework for Surface-wave Inversion}

\subsection{Transformer-based Architecture for Dispersion Curve Inversion}

\begin{figure}[!ht]
        \centering
        \includegraphics[width=0.85\textwidth]{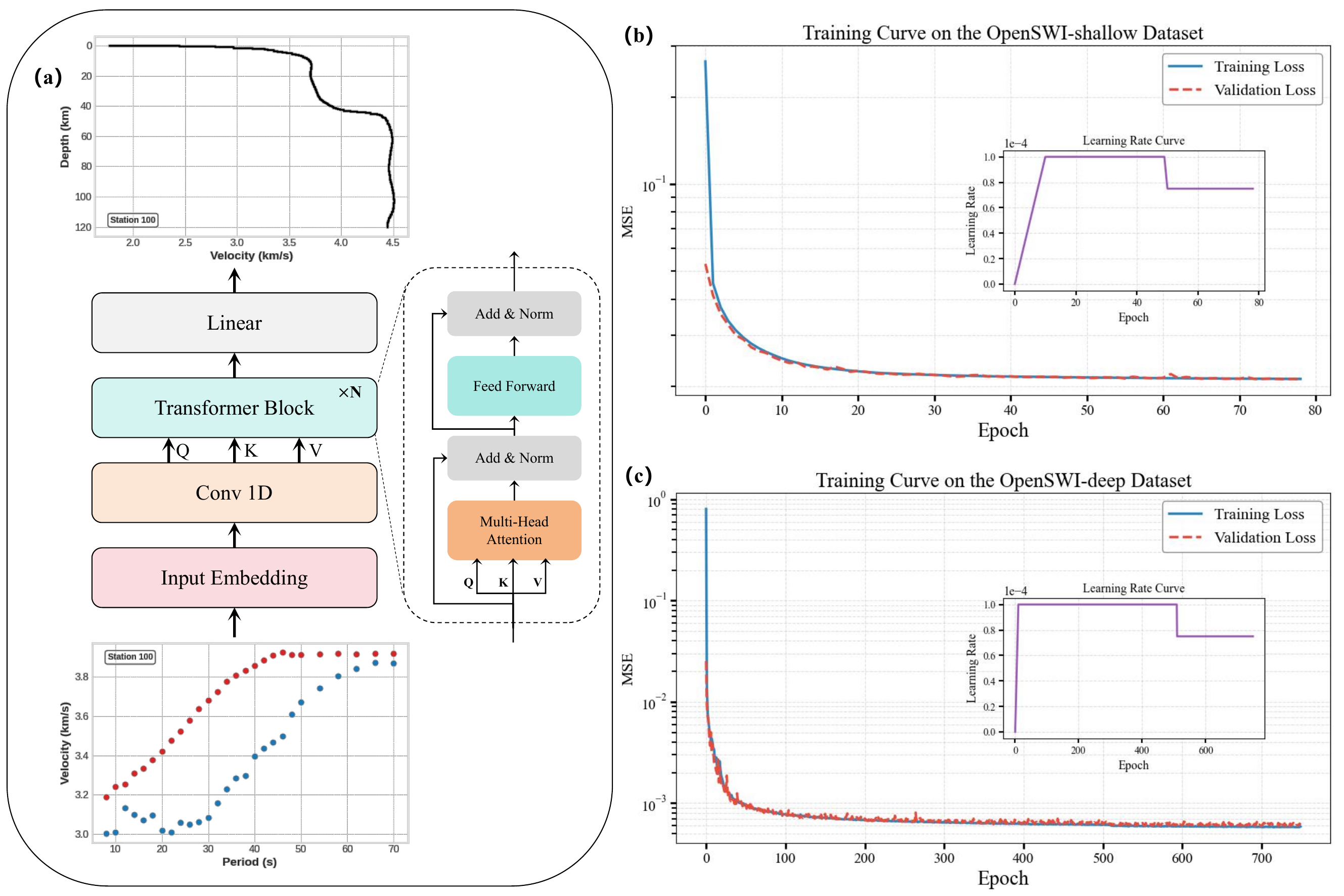}
        \caption{(a) The architecture of the deep neural network (Transformer) used in this work for surface wave dispersion curve inversion. The Training (blue) and validation (red) loss curve on the (b) OpenSWI-shallow, and (c) OpenSWI-deep datasets. The learning rate curve are presents in the inner figure with purple line.}
        \label{fig10:dispformer_network}
    \end{figure}

    Deep learning-based surface-wave dispersion curve inversion seeks to learn a nonlinear mapping from input dispersion curves (including period, phase velocity, and group velocity) to corresponding 1-D subsurface shear-wave velocity profiles. In this study, we adopt a widely used Transformer-based architecture (Figure~\ref{fig10:dispformer_network}a) to enable end-to-end inversion \citep{liu_2025_DispFormer, jiang_2025_OneFitAll}. The input to the model is a $3 \times N$ dispersion curve matrix, where the three rows represent period, phase velocity, and group velocity, and $N$ denotes the number of sampled points. The model initially embeds the input via three separate $1 \times 1$ convolutional neural network (CNN) layers, yielding a feature representation of size $3 \times N \times E$, where $E$ is the feature dimension. These embedded features are then processed through multiple Transformer blocks, which employ self-attention mechanisms to capture long-range dependencies across the dispersion curves. This global context modeling enhances the stability and accuracy of inversion results. Finally, a feature projection layer maps the global features extracted by the Transformer to a velocity profile of length $M$, where $M$ corresponds to the number of target depth layers, producing the final inversion output.

    Given that the period range and target depth in real observational data vary, and that the maximum inversion depth strongly correlates with the observed period range, we incorporate the depth-aware strategy proposed by \citet{liu_2025_DispFormer} during training. This approach dynamically computes the maximum wavelength (period multiplied by velocity) for each input and adaptively determines the effective output depth range, thereby suppressing predictions at irrelevant depths and improving inversion accuracy. For the loss function, we adopt the Mean Squared Error (MSE), calculated exclusively over the effective depth range between predicted and ground-truth velocity profiles. To enhance robustness against noise and missing data commonly encountered in practice, we simulate these effects during data loading by adding 3\% random Gaussian noise and randomly masking 10\% of the dispersion data points.

    Regarding training configuration, we use a larger batch size of 2048 and limit training to 100 epochs for the shallow dispersion dataset (OpenSWI-shallow) to optimize large-scale training efficiency. For the deeper dataset (OpenSWI-deep), a smaller batch size of 512 and up to 1000 epochs are employed. To avoid overfitting and reduce unnecessary computation, we adopt an early stopping strategy, terminating training when the validation loss does not improve for 30 consecutive epochs for OpenSWI-shallow and 50 epochs for OpenSWI-deep. Both datasets are trained using the Adam optimizer, combined with a learning rate scheduler that integrates warm-up and step decay strategies to enhance training stability. During warm-up, the learning rate increases linearly from $1 \times 10^{-9}$ to $1 \times 10^{-4}$ over approximately 2 epochs for OpenSWI-shallow and 10 epochs for OpenSWI-deep. In the subsequent step decay phase, the learning rate is reduced to 75\% of its value every 20 epochs for OpenSWI-shallow and every 200 epochs for OpenSWI-deep. Figures~\ref{fig10:dispformer_network}b and \ref{fig10:dispformer_network}c present the training and validation error curves for both datasets alongside their corresponding learning rate schedules.

    Model performance is first evaluated on the test sets by comparing predicted and ground-truth velocity profiles using Root Mean Squared Error (RMSE). Beyond this quantitative validation, the trained models are applied to real observational data to assess their generalization capabilities. Instead, we compare the synthetic and observed dispersion curves to compute the misfit errors, and assess the inversion quality by analyzing the distribution of these errors, including their mean and variance.

\subsection{Large-scale Training with the OpenSWI-shallow and OpenSWI-deep Datasets}

    To comprehensively assess the effectiveness of the proposed deep neural network model for surface wave dispersion curve inversion, we conducted systematic training on both the OpenSWI-shallow and OpenSWI-deep datasets. Detailed architectural hyperparameters are provided in Appendix~\ref{sec:appendix_network_structure_details}. To ensure balanced representation across the training, validation, and test subsets, we employed stratified sampling strategies. Specifically, for the OpenSWI-shallow dataset, stratification was based on geological structure types (Flat, Flat-Fault, Fold, Fold-Fault, and Field), using a 90\%/5\%/5\% split. For the OpenSWI-deep dataset, stratification was performed by geographic regions of the source models, following the same partitioning ratio.

    During training, both training and validation errors were continuously monitored, as illustrated in Figure~\ref{fig10:dispformer_network}b and c. For both datasets, the error curves demonstrate stable convergence, suggesting that the model effectively captures the nonlinear relationship between surface wave dispersion curves and subsurface shear-wave velocity profiles. After training, evaluation on the held-out test sets yielded RMSE values of 0.1467 km/s for OpenSWI-shallow and 0.048 km/s for OpenSWI-deep, indicating that the predicted velocity models closely match the ground-truth profiles and confirming the model’s high inversion accuracy under varying geological conditions.
    
\begin{figure}[!ht]
        \centering
        \includegraphics[width=1.0\textwidth]{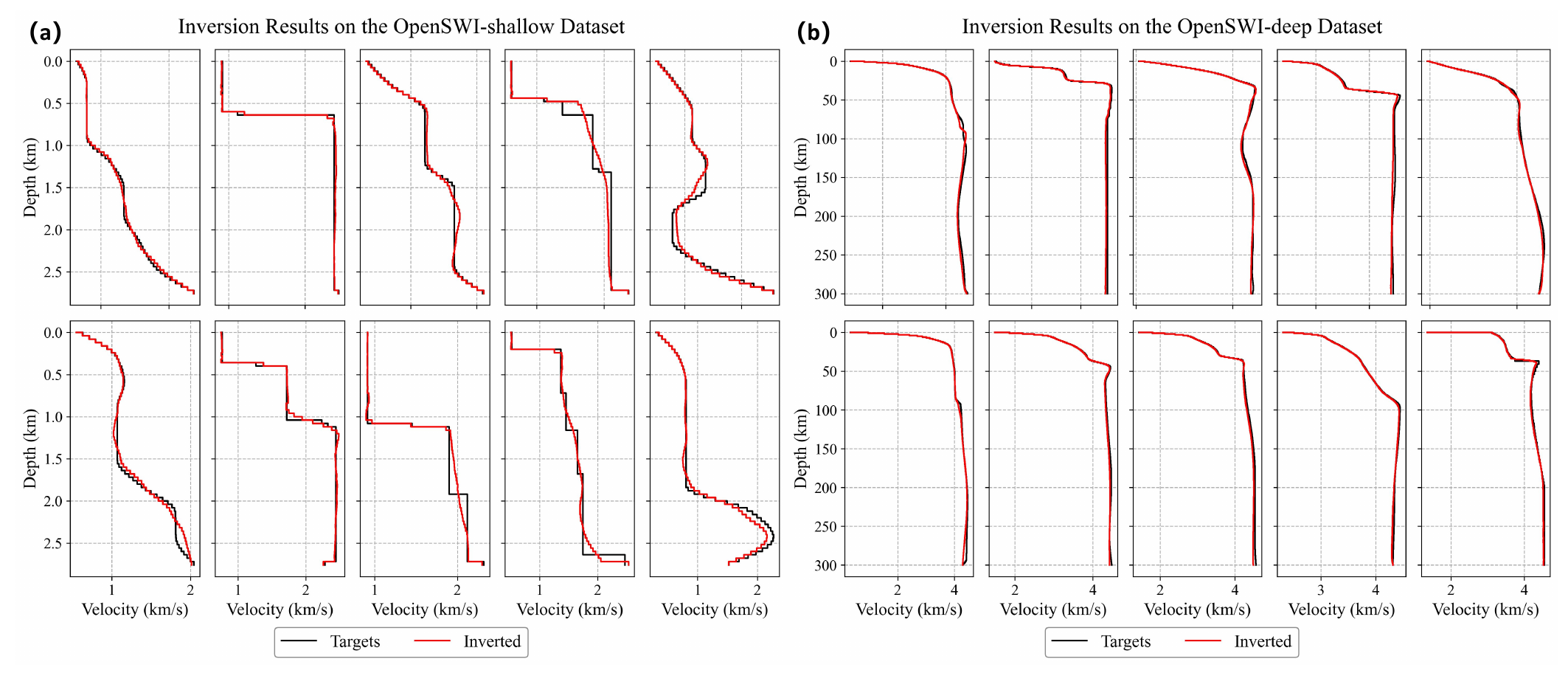}
        \caption{Representative inversion results on the test subsets of (a) OpenSWI-shallow and (b) OpenSWI-deep. The black lines represent the ground-truth velocity profiles, while the red lines denote the predicted results obtained by the trained neural network.}
        \label{fig11:openswi_deep_results}
    \end{figure}

    Representative inversion results from both datasets are shown in Figures~\ref{fig11:openswi_deep_results}a and \ref{fig11:openswi_deep_results}b, further demonstrating the model’s capability to reconstruct subsurface velocity structures with high fidelity, including at greater depths. These results validate the generalization ability and practical applicability of the proposed method across diverse geological settings. It is worth noting that OpenSWI-shallow includes significantly more samples than OpenSWI-deep and encompasses a wider variety of geologically diverse and structurally complex velocity models. Consequently, achieving optimal performance on this dataset demands more specialized architectural designs and training strategies. While the model maintains strong overall inversion quality, it tends to oversmooth regions characterized by strong heterogeneity or abrupt structural changes, resulting in slightly muted responses in complex geological zones. This smoothing effect highlights current limitations in resolving fine-scale structural features and underscores the need for future enhancements, such as structure-aware regularization or multi-scale modeling techniques, to improve the representation of intricate subsurface variations.

\subsection{Generalization Testing on Real-world Observations Using OpenSWI-real}

    To evaluate the generalization capability of deep neural networks trained entirely on synthetic data, we directly applied the pretrained models to the OpenSWI-real dataset, which includes two representative real-world regions: Long Beach (shallow) and CSRM (deep).

\begin{figure}[!ht]
        \centering
        \includegraphics[width=1.0\textwidth]{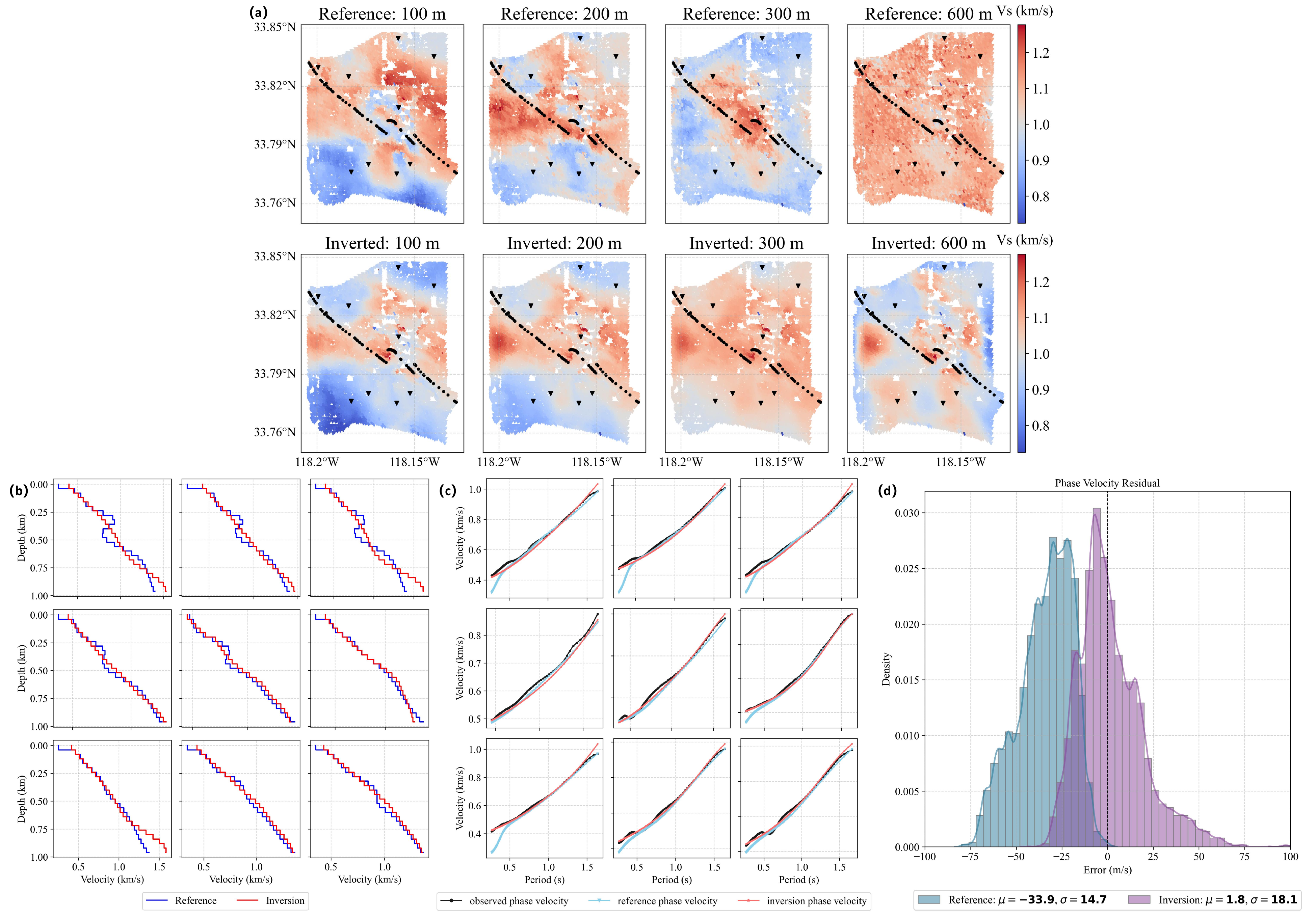}
        \caption{Generalization performance on real-world Long Beach data from the OpenSWI-real dataset. (a) Reference \citep{fu_2022_Improved} and predicted shear-wave velocity slices at depths of 100, 200, 300, and 600~m. (b) 1-D shear-wave velocity profiles at nine representative locations, with reference and predicted models shown in blue and red, respectively. (c) Comparison of phase velocity dispersion curves, including observed curves (black), synthetic curves from the reference model (blue) and the predicted model (red). (d) Error distributions of phase velocity with respect to observed curves, based on synthetic dispersion curves from the reference (blue) and predicted (purple) models.}
        \label{fig12:openswi_shallow_longbeach}
    \end{figure}

    In the shallow case, we used phase velocity dispersion curves from 5,297 stations in the Long Beach area as input to the shallow inversion network. The model generated a 1-D shear-wave velocity profile for each station, which were then assembled into a 3-D velocity model of the region. Figure~\ref{fig12:openswi_shallow_longbeach}a presents horizontal slices of the predicted model at depths of 100~m, 200~m, 400~m, and 600~m, alongside the corresponding reference model. Figure~\ref{fig12:openswi_shallow_longbeach}b compares selected 1-D profiles from both models. Notably, despite the complete absence of Long Beach data during training, the model successfully reconstructs key subsurface velocity structures. In particular, the predicted profiles at 100~m and 200~m show excellent agreement with the reference model. Figure~\ref{fig12:openswi_shallow_longbeach}c shows the observed dispersion curves (black), as well as synthetic curves generated from the reference model (blue) and the neural network predictions (red). To quantitatively evaluate inversion performance, we computed the misfit between observed dispersion curves and those derived from the predicted velocity profiles. Figure~\ref{fig12:openswi_shallow_longbeach}d summarizes the error distributions. For the reference model, the mean and variance of the misfit are –33.9m/s and 14.7(m/s)\textsuperscript{2}, respectively, whereas the neural network predictions yield a mean misfit of 1.8m/s and a variance of 18.1(m/s)\textsuperscript{2}. These results demonstrate that the pretrained model generalizes effectively to real observational data and, in many cases, even outperforms the reference model, particularly in shallow geological settings.

\begin{figure}[!ht]
        \centering
        \includegraphics[width=0.83\textwidth]{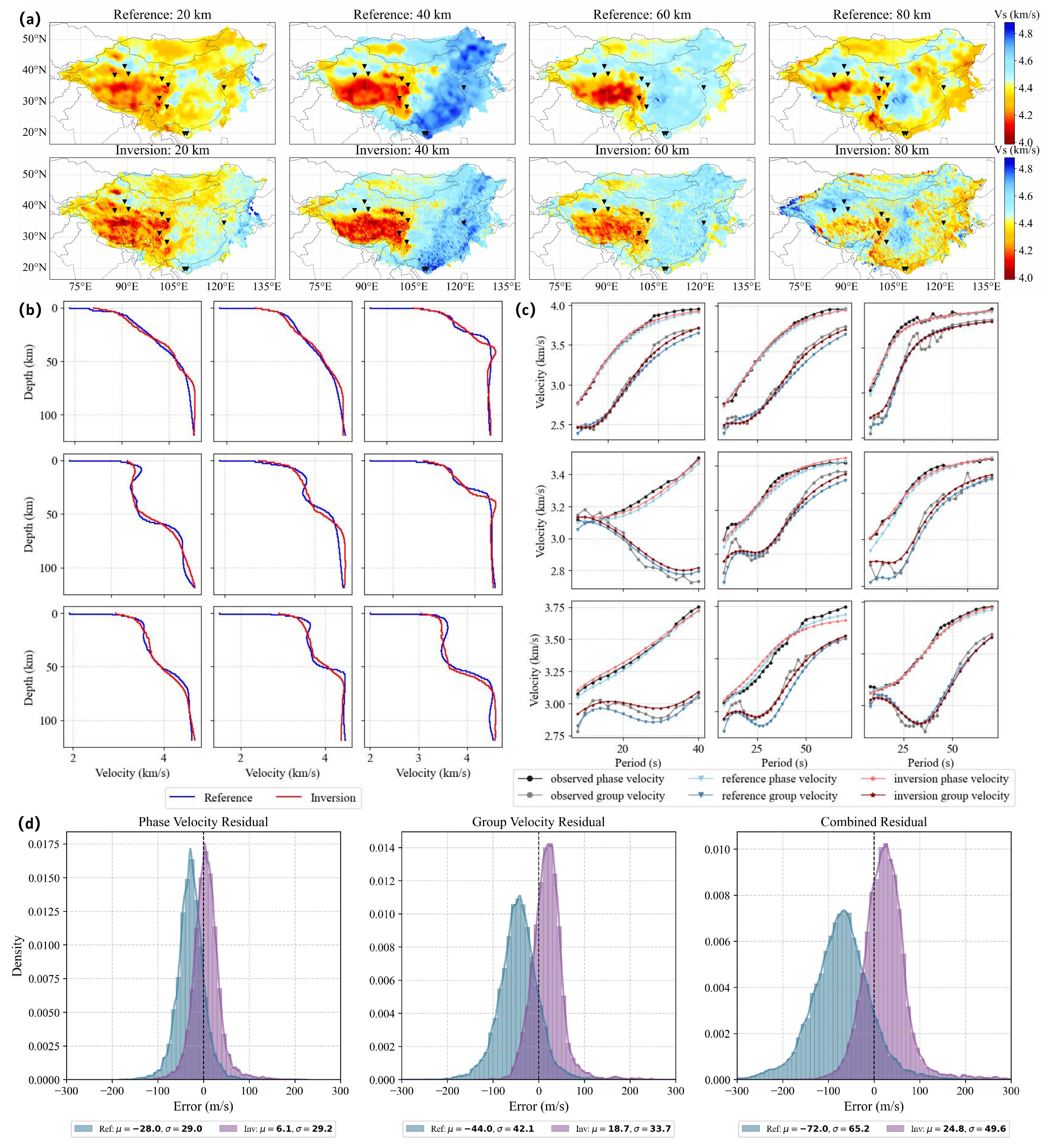}
        \caption{Generalization performance on real-world CSRM data from the OpenSWI-real dataset. (a) Reference \citep{xiao_2024_CSRM10} and predicted $v_s$ slices at depths of 20, 40, 60, and 80~km. (b) 1-D shear-wave velocity profiles at nine representative locations, with reference and predicted models shown in blue and red, respectively. (c) Comparison of phase and group velocity dispersion curves, including observed curves (black), synthetic curves from the reference model (blue for group velocity, light blue for phase velocity), and from the predicted model (red for group velocity, pink for phase velocity). (d) Error distributions of phase velocity (left), group velocity (middle), and their sum (right) with respect to observed curves, based on synthetic dispersion curves from the reference (blue) and predicted (purple) models.}
        \label{fig13:openswi_deep_csrm}
    \end{figure}

    For the deep case, we applied the pretrained deep inversion network to both phase and group velocity dispersion curves at 12,901 grid points provided by the CSRM project \citep{wen_2023_China,xiao_2024_CSRM10}. Figure~\ref{fig13:openswi_deep_csrm}a compares the predicted and reference velocity structures at depths of 20~km, 40~km, 60~km, and 80~km. Figure~\ref{fig13:openswi_deep_csrm}b shows 1-D profile comparisons at nine representative grid points, where black lines denote the reference models and red lines indicate the neural network predictions. Figure~\ref{fig13:openswi_deep_csrm}c presents the observed dispersion curves (black), along with synthetic curves generated from the reference model (blue) and the predicted models (red). Figure~\ref{fig13:openswi_deep_csrm}d displays the distribution of misfits between synthetic and observed dispersion curves across all grid points. The reference model achieves a mean misfit of –72.9m/s with a variance of 65.5(m/s)\textsuperscript{2}, while the neural network results exhibit a mean misfit of 24.8m/s and a lower variance of 49.6(m/s)\textsuperscript{2}. These findings suggest that the trained network can recover deep crustal velocity structures with accuracy comparable to, or better than, that of the reference model—even without any fine-tuning on real data.
    
    In summary, these experiments confirm the strong generalization ability of the proposed method across a broad range of geological settings and depth regimes. More importantly, they highlight the effectiveness of the OpenSWI dataset series in enabling the training and evaluation of deep learning-based inversion techniques. With its extensive geological diversity, structural complexity, and broad spatial coverage, the OpenSWI dataset provides a solid foundation for learning transferable representations. As demonstrated, the resulting models can produce high-quality inversion results on real-world observations without retraining or domain adaptation, positioning OpenSWI as a valuable benchmark for advancing deep learning in realistic geophysical applications.

%
%
\section{Discussion}

    The OpenSWI dataset marks a substantial advancement in the development of AI-ready benchmark datasets for surface wave dispersion curve inversion. Compared to existing public datasets, OpenSWI offers significantly larger scale, broader spatial coverage, and enhanced geological diversity. Specifically, the OpenSWI-shallow subset contains over 22 million 1-D velocity profiles and their associated dispersion curves representing shallow subsurface structures (depths < 3 km), while the OpenSWI-deep subset comprises approximately 1.28 million samples covering deeper Earth structures down to 300 km. In addition, the OpenSWI-real dataset provides real-world observational data for validating inversion methods under practical conditions. This comprehensive suite enables robust evaluation of machine learning–based approaches across synthetic and real data scenarios. Furthermore, a complete dataset construction toolkit, SWIDP, is released alongside the dataset, allowing users to flexibly generate customized datasets tailored to specific research needs.

    Experimental results show that deep learning models trained exclusively on synthetic data from OpenSWI exhibit strong generalization to real-world observations, even without fine-tuning. This underscores the importance of large-scale, high-fidelity synthetic datasets in overcoming the challenges posed by the limited availability and annotation complexity of real seismic data. Practically, this indicates that reliable inversion results can be obtained even in regions with sparse or low-quality observations, thereby lowering the threshold for deploying machine learning models in real-world geophysical applications.
    
    Despite these strengths, several limitations remain. First, the derivation of $v_p$ and $\rho$ from $v_s$ through empirical relationships may introduce systematic biases, especially in regions with complex or atypical geological structures \citep{brocher_2005_Empirical}. Second, although OpenSWI spans a wide array of tectonic and geological environments, it still underrepresents certain extreme (e.g. anisotropic media, fluid-saturated layers) or geodynamically active (e.g. mid-ocean ridges, highly deformed orogenic belts) settings, limiting its applicability in those areas. Third, the current dataset focuses primarily on fundamental-mode Rayleigh wave dispersion curves and does not incorporate higher modes or additional geophysical observables (e.g., ellipticity, receiver functions), which constrains its utility for joint inversion frameworks \citep{liu_2024_Multimodal, jiang_2025_OneFitAll}. Lastly, although OpenSWI incorporates a degree of noise and data incompleteness, it does not fully capture the complexities of real-world measurements, including uncertainties in source characteristics, instrument responses, and acquisition-related biases.

    Future developments can be pursued along several interrelated directions. First, expanding the dataset’s geographic coverage and geological diversity, particularly in tectonically extreme regions, would broaden its applicability. Second, integrating data across different modes, period ranges, and geological settings could enable more robust inversion approaches and improve transferability across regions. Third, incorporating additional real observational data to construct datasets suitable for hybrid or transfer learning would further enhance model generalization in field applications. Finally, including higher-mode dispersion curves and complementary geophysical observables would support more comprehensive multi-modal and multi-physics inversion strategies. We envision OpenSWI as a long-term, evolving community resource that will continue to drive data-driven advances in surface wave inversion and geophysical imaging.

%
%
\conclusions
    
    In this study, we present OpenSWI, the first AI-ready benchmark dataset at the tens-of-millions scale specifically designed for surface wave dispersion curve inversion, along with a complete data generation toolkit, SWIDP. The dataset encompasses both shallow and deep subsurface velocity structures across a wide range of geological settings. Its large scale, geological diversity, and standardized formats for velocity profiles and dispersion curves provide a robust foundation for evaluating machine learning–based inversion methods. Experimental results show that models trained entirely on synthetic data from OpenSWI can generalize effectively to real-world observations, highlighting the dataset's practical value in improving the robustness and applicability of data-driven inversion approaches. Future developments will focus on expanding the dataset's geographic and geological coverage, incorporating additional geophysical observables to support more complex joint inversion tasks, and explore deeper integration with real observational data. We expect OpenSWI to serve as an open, continuously evolving community resource that promotes reproducible research and supports the broader application of machine learning methods in geophysical imaging.

%
%

\codedataavailability{
    All codes, datasets, and experimental results used in this study have been made publicly available to facilitate reproducibility, validation, and further development by the research and industrial communities. The Python toolkit \texttt{SWIDP}, has been released on the GitHub repository (\url{https://github.com/liufeng2317/OpenSWI}). This toolkit includes modules for 1-D velocity profile extraction and augmentation, layer parameter conversion, dispersion curve computation, and 2-D velocity model augmentation based on diffusion models. The \texttt{OpenSWI} dataset, including \texttt{OpenSWI-shallow}, \texttt{OpenSWI-deep}, and \texttt{OpenSWI-real}, is provided in a unified format with complete metadata, and can be accessed from \url{https://huggingface.co/datasets/LiuFeng2317/OpenSWI}. The deep learning training codes, pretrained model weights, and key experimental results used in this work are also openly accessible to support future research and method development.
}
    
\clearpage

\appendix

\section{Illustrative Code Examples for the OpenSWI-shallow Generation Workflow with SWIDP}
\label{sec:appendix_swidp_shallow}
\begin{lstlisting}[]
# Import all core functions from the SWIDP package
from SWIDP import *
# initialize the model
model = SWIModel()
# Step 1: Data extraction and duplication removal
# 1-1: Load velocity model
model.load_openfwi_velocity_model()
# 1-2: Extract velocity profiles
model.get_velocity_profiles()
# 1-3: Convert units (e.g., m/s to km/s)
model.convert_unit()       
# 1-4: Remove duplicates
model.unique_profiles()                
# 1-5: Convert vp to vs
model.transform_vp_to_vs()                   
# Step 2: Data augmentation
# 2-1: Merge adjacent layers with similar vs
model.combine_same_vs()                
# 2-2: Remove thin layers
model.remove_thin_layer()             
for i in range(augment_times):
    # 2-3: Perturb velocity and layer thickness
    model.perturb_vs_depth()          
    # 2-4: Interpolate to original depth
    model.interpolate_profile()     
    # 2-5: Generate full velocity model
    model.transform_vs_to_vel_model()
# Step 3: Compute dispersion curves
# 3-1: Generate period samples
model.generate_mixed_samples()  # uniform, random, and  logarithmic sampling
# 3-2: Calculate dispersion curves
model.calculate_dispersion()
# Step 4: Save data
model.save_velocity_model()     # [depth, vp, vs, density]
model.save_dispersion_curves()  # [period, phase velocity, group velocity]
\end{lstlisting}
\clearpage

\section{Illustrative Code Examples for the OpenSWI-deep Generation Workflow with SWIDP}    
\label{sec:appendix_swidp_deep}
\begin{lstlisting}[]
# Import all core functions from the SWIDP package
from SWIDP import *
# initialize the model
model = SWIModel()
# 1-1 load velocity profiles
model.extract_velocity_profiles()
# 1-2 interpolate velocity profiles
model.interpolate_velocity_profiles()
# 1-3 combine thin layers (remove extremely thin layers)
model.combine_thin_sandwich()
# 1-4 smooth velocity profiles (optional)
model.smooth_vs_by_node_interp()
# 2. find moho depth
model.find_moho_depth()
# 3-1 augment velocity model (Crust-Moho-Mantle)
model.augment_crust_moho_mantle()
# 3-2 transform velocity model to velocity model
model.transform_vs_to_vel_model() 
# 4-1 generate mixed samples
model.generate_mixed_samples() # uniform, random, and  logarithmic sampling
# 4-2 calculate dispersion curves
model.calculate_dispersion()
# 5. save velocity model and dispersion curves
model.save_velocity_model() # [depth, vp, vs, density]
model.save_dispersion_curves() # [period, phase velocity, group velocity]
\end{lstlisting}
\clearpage

\section{Diffusion Probabilistic Models for Continually Augmenting the OpenSWI-shallow Subsets}
\label{sec:appendix_ddpm}

\subsection{Introduction to Denoising Diffusion Probabilistic Models (DDPMs)}

    Denoising Diffusion Probabilistic Models (DDPMs) are a class of powerful generative models that progressively refine noisy data to generate realistic outputs \citep{ho_2020_Denoising, taufik_2024_Learned}. The core principle of DDPMs involves a two-step diffusion process: a forward process in which noise is progressively added to the data, and a reverse process in which the model learns to remove the noise and recover the original data distribution. In this study, DDPMs are applied to model and augment geological structures within the OpenFWI dataset \citep{deng_2021_OpenFWI}, which consists of five subsets: FlatVel-A, FlatFault-A, CurveVel-A, CurveFault-A, and Style-A. These subsets represent various subsurface geophysical features, and by learning their distribution characteristics, DDPMs are capable of generating new, physically plausible velocity models that exhibit complex geological features such as faults, folds, and field-style structures.

\subsection{Core Principle of DDPM}

    DDPMs are based on two main processes:
    \begin{itemize}
        \item \textbf{Forward diffusion}: Starting from an input data point, Gaussian noise is progressively added in multiple steps, transforming the data into pure noise.
        \item \textbf{Reverse diffusion}: The model learns to reverse this process, starting from random noise and progressively denoising it to recover the underlying data distribution.
    \end{itemize}
    
    The reverse denoising process is learned by training a neural network to predict the noise added at each diffusion step. The objective is to minimize the difference between the predicted noise and the actual noise, enabling the model to generate realistic data that follows the original distribution. Formally, the training loss function is defined as:
    \begin{align}
        L(\theta) = \mathbb{E}_{q(\mathbf{x}_0)} \left[ \sum_{t=1}^{T} \| \epsilon_\theta(\mathbf{x}_t, t) - \epsilon_t \|^2 \right]
    \end{align}
    where \(\epsilon_\theta\) is the predicted noise at step \(t\), and \(\epsilon_t\) is the actual noise added during the forward diffusion process.
    
    For further details on the DDPM methodology, please refer to \citet{ho_2020_Denoising}. Additionally, an implementation of DDPM in PyTorch is available at \url{https://github.com/lucidrains/denoising-diffusion-pytorch}.

\subsection{Model Architecture and Training Configuration}

    The DDPM model used in this study follows a U-Net architecture with the following key components:
    
    \begin{itemize}
        \item \textbf{U-Net architecture}: A convolutional neural network with an encoder-decoder structure. The encoder reduces the spatial resolution, and the decoder restores it to the original resolution ($64 \times 64$). The architecture includes residual blocks and batch normalization.
        \item \textbf{Noise schedule}: A linear noise schedule is applied during the forward diffusion process, where the variance of the Gaussian noise increases progressively with each step (total 1000 steps).
        \item \textbf{Optimizer}: Adam optimizer with a learning rate of \(1e^{-6}\), \(\beta_1 = 0.9\), and \(\beta_2 = 0.999\).
        \item \textbf{Training duration}: The model was trained for 5000 epochs with a batch size of 256.
    \end{itemize}
    
    The training objective is to minimize the difference between the predicted and actual noise added during the forward diffusion process, as described by the loss function in the previous section.

\subsection{DDPM sampling and OpenSWI-shallow datasets Generation}

    After training, the DDPM model is used for continuous data augmentation by generating new velocity models. This process involves sampling Gaussian noise and running the reverse diffusion process to produce realistic velocity models. The generated models reflect a variety of subsurface features, such as faults and complex sedimentary structures, ensuring physical plausibility.
    
    To facilitate the integration of the DDPM-generated models into the OpenSWI-shallow dataset, we provide a set of tools in the SWIDP pipline. These tools enable the extraction and conversion of the DDPM sampling results into 1-D velocity models, as required by OpenSWI-shallow. The process includes the following key steps:
    \begin{itemize}
        \item \textbf{DDPM sampling}: The DDPM model generates new velocity models by progressively denoising random Gaussian noise.
        \item \textbf{Denormalization}: The generated models, initially in normalized form, are denormalized to match the required velocity range.
        \item \textbf{Profile extraction and rationalization}: The velocity models are then extracted into 1-D velocity profiles and rationalized to ensure geological consistency.
        \item \textbf{Dispersion curve calculation}: The rationalized 1-D velocity profiles are used to calculate the corresponding dispersion curves, which are essential for surface wave inversion tasks.
    \end{itemize}
    
    By continually generating new data and performing the above operations, the OpenSWI-shallow dataset is augmented with a diverse set of realistic velocity profiles, further expanding the dataset's coverage and variability for improved inversion model robustness.




    
\clearpage

\section{Transformer-based Network Architecture for Different Datasets}
\label{sec:appendix_network_structure_details}

\begin{table}[ht]
    \centering
    \caption{Transformer-based Network Architecture for Different Datasets}
    \begin{tabular}{l|c|c|c|c|c}
    \hline
    \textbf{Dataset} & \textbf{Input Shape} & \textbf{Embedding Dim.} & \textbf{Transformer Blocks} & \textbf{Attention Heads} & \textbf{Output Shape} \\ \hline
    OpenSWI-shallow   & 3 $\times$ 100        & 64                      & 3                          & 8                        & 1 $\times$ 70         \\
    OpenSWI-deep      & 3 $\times$ 300        & 128                     & 3                          & 8                        & 1 $\times$ 300        \\ \hline
    \end{tabular}
    \vspace{0.2cm}
    
    \footnotesize{
        \textit{Note}: The input shape consists of three features: period, phase velocity, and group velocity. The output shape corresponds to the shear-wave velocity (\(v_s\)).
    }
\end{table}

\noappendix       






\clearpage
\authorcontribution{
F. LIU: Conceptualization, Data Curation, Methodology, Software, Formal Analysis, Writing - Original Draft, Writing - Review \& Editing, Visualization \\
S. ZHAO: Conceptualization, Data Curation, Methodology, Formal Analysis, Writing - Review \& Editing \\
X. GU: Resources, Supervision, Writing - Review \& Editing, Visualization \\
F. LING: Resources, Supervision, Writing - Review \& Editing \\
P. ZHUANG: Resources, Supervision, Writing - Review \& Editing \\
Y. LI: Supervision, Writing - Review \& Editing, Project Administration \\
R. SU: Supervision, Writing - Review \& Editing, Funding Acquisition \\
L. Fang: Writing - Review \& Editing \\
L. Zhou: Writing - Review \& Editing \\
J. HUANG: Writing - Review \& Editing \\
L. BAI: Supervision, Writing - Review \& Editing, Funding Acquisition 
} 

\competinginterests{The authors declare no competing interests.} 


\begin{acknowledgements}
The authors thanks the Shanghai Artificial Intelligence Laboratory for providing computational resources for this project. This research is supported by the the Shanghai Municipal Science and Technology Major Project.
\end{acknowledgements}



\bibliographystyle{copernicus}
\bibliography{reference.bib}

\begin{thebibliography}{63}
\providecommand{\natexlab}[1]{#1}
\providecommand{\url}[1]{\texttt{#1}}
\providecommand{\urlprefix}{}
\expandafter\ifx\csname urlstyle\endcsname\relax
  \providecommand{\doi}[1]{https://doi.org/\discretionary{}{}{}#1}\else
  \providecommand{\doi}{https://doi.org/\discretionary{}{}{}\begingroup \urlstyle{rm}\Url}\fi

\bibitem[{Aleardi and Stucchi(2021)}]{aleardi_2021_Hybrid}
Aleardi, M. and Stucchi, E.: A Hybrid Residual Neural Network--{{Monte Carlo}} Approach to Invert Surface Wave Dispersion Data, Near Surface Geophysics, 19, 397--414, \doi{10.1002/nsg.12163}, 2021.

\bibitem[{Bensen et~al.(2007)Bensen, Ritzwoller, Barmin, Levshin, Lin, Moschetti, Shapiro, and Yang}]{bensen_2007_Processing}
Bensen, G.~D., Ritzwoller, M.~H., Barmin, M.~P., Levshin, A.~L., Lin, F., Moschetti, M.~P., Shapiro, N.~M., and Yang, Y.: Processing Seismic Ambient Noise Data to Obtain Reliable Broad-Band Surface Wave Dispersion Measurements, Geophysical Journal International, 169, 1239--1260, \doi{10.1111/j.1365-246X.2007.03374.x}, 2007.

\bibitem[{Berg et~al.(2020)Berg, Lin, Allam, Schulte-Pelkum, Ward, and Shen}]{berg_2020_Shear}
Berg, E.~M., Lin, F.-C., Allam, A., Schulte-Pelkum, V., Ward, K.~M., and Shen, W.: Shear Velocity Model of Alaska via Joint Inversion of Rayleigh Wave Ellipticity, Phase Velocities, and Receiver Functions across the Alaska Transportable Array, Journal of Geophysical Research: Solid Earth, 125, \doi{10.1029/2019jb018582}, 2020.

\bibitem[{Blom et~al.(2020)Blom, Gokhberg, and Fichtner}]{blom_2020_Seismic}
Blom, N., Gokhberg, A., and Fichtner, A.: Seismic Waveform Tomography of the Central and Eastern {{Mediterranean}} Upper Mantle, Solid Earth, 11, 669--690, \doi{10.5194/se-11-669-2020}, 2020.

\bibitem[{Brocher(2005)}]{brocher_2005_Empirical}
Brocher, T.~M.: Empirical Relations between Elastic Wavespeeds and Density in the Earth's Crust, Bulletin of the Seismological Society of America, 95, 2081--2092, \doi{10.1785/0120050077}, 2005.

\bibitem[{Cai et~al.(2022)Cai, Qiu, and Niu}]{cai_2022_SemiSupervised}
Cai, A., Qiu, H., and Niu, F.: Semi-{{Supervised Surface Wave Tomography With Wasserstein Cycle}}-{{Consistent GAN}}: {{Method}} and {{Application}} to {{Southern California Plate Boundary Region}}, Journal of Geophysical Research: Solid Earth, 127, e2021JB023\,598, \doi{10.1029/2021JB023598}, 2022.

\bibitem[{Cao et~al.(2020)Cao, Earp, De~Ridder, Curtis, and Galetti}]{cao_2020_Nearrealtime}
Cao, R., Earp, S., De~Ridder, S. A.~L., Curtis, A., and Galetti, E.: Near-Real-Time near-Surface {{3D}} Seismic Velocity and Uncertainty Models by Wavefield Gradiometry and Neural Network Inversion of Ambient Seismic Noise, GEOPHYSICS, 85, KS13--KS27, \doi{10.1190/geo2018-0562.1}, 2020.

\bibitem[{Chen et~al.(2025)Chen, Xia, Feng, Cheng, Pang, and Hong}]{chen_2025_Why}
Chen, X., Xia, J., Feng, J., Cheng, F., Pang, J., and Hong, Y.: Why {{Choose Deep Learning}} for {{Surface-Wave Inversion}}, Surveys in Geophysics, 46, 695--722, \doi{10.1007/s10712-025-09882-y}, 2025.

\bibitem[{Colli et~al.(2013)Colli, Fichtner, and Bunge}]{colli_2013_Full}
Colli, L., Fichtner, A., and Bunge, H.-P.: Full Waveform Tomography of the Upper Mantle in the South Atlantic Region: Imaging a Westward Fluxing Shallow Asthenosphere?, Tectonophysics, 604, 26--40, \doi{10.1016/j.tecto.2013.06.015}, 2013.

\bibitem[{{\c C}ubuk-Sabuncu et~al.(2017){\c C}ubuk-Sabuncu, Taymaz, and Fichtner}]{sabuncu_2017_3D}
{\c C}ubuk-Sabuncu, Y., Taymaz, T., and Fichtner, A.: 3-{{D}} Crustal Velocity Structure of Western {{Turkey}}: Constraints from Full-Waveform Tomography, Physics of the Earth and Planetary Interiors, 270, 90--112, \doi{10.1016/j.pepi.2017.06.014}, 2017.

\bibitem[{Deng et~al.(2021)Deng, Feng, Wang, Zhang, Jin, Feng, Zeng, Chen, and Lin}]{deng_2021_OpenFWI}
Deng, C., Feng, S., Wang, H., Zhang, X., Jin, P., Feng, Y., Zeng, Q., Chen, Y., and Lin, Y.: {{OpenFWI}}: {{Large-scale}} Multi-Structural Benchmark Datasets for Full Waveform Inversion, Neural Information Processing Systems, 2021.

\bibitem[{Feng et~al.(2023)Feng, Wang, Deng, Feng, Liu, Zhu, Jin, Chen, and Lin}]{feng_2023_EFWI}
Feng, S., Wang, H., Deng, C., Feng, Y., Liu, Y., Zhu, M., Jin, P., Chen, Y., and Lin, Y.: {{EFWI}} Multiparameter Benchmark Datasets for Elastic Full Waveform Inversion of Geophysical Properties, Advances in Neural Information Processing Systems, 36, 23\,701--23\,713, 2023.

\bibitem[{Fichtner and Villase{\~n}or(2015)}]{fichtner_2015_Crust}
Fichtner, A. and Villase{\~n}or, A.: Crust and Upper Mantle of the Western {{Mediterranean}} -- Constraints from Full-Waveform Inversion, Earth and Planetary Science Letters, 428, 52--62, \doi{10.1016/j.epsl.2015.07.038}, 2015.

\bibitem[{Fichtner et~al.(2006)Fichtner, Bunge, and Igel}]{fichtner_2006_Adjoint}
Fichtner, A., Bunge, H.-P., and Igel, H.: The Adjoint Method in Seismology, Physics of the Earth and Planetary Interiors, 157, 86--104, \doi{10.1016/j.pepi.2006.03.016}, 2006.

\bibitem[{Fichtner et~al.(2009)Fichtner, Kennett, Igel, and Bunge}]{fichtner_2009_Full}
Fichtner, A., Kennett, B. L.~N., Igel, H., and Bunge, H.-P.: Full Seismic Waveform Tomography for Upper-Mantle Structure in the Australasian Region Using Adjoint Methods, Geophysical Journal International, 179, 1703--1725, \doi{10.1111/j.1365-246x.2009.04368.x}, 2009.

\bibitem[{Fichtner et~al.(2010)Fichtner, Kennett, Igel, and Bunge}]{fichtner_2010_Full}
Fichtner, A., Kennett, B.~L., Igel, H., and Bunge, H.-P.: Full Waveform Tomography for Radially Anisotropic Structure: New Insights into Present and Past States of the Australasian Upper Mantle, Earth and Planetary Science Letters, 290, 270--280, \doi{10.1016/j.epsl.2009.12.003}, 2010.

\bibitem[{Fichtner et~al.(2013)Fichtner, Trampert, Cupillard, Saygin, Taymaz, Capdeville, and Villase{\~n}or}]{fichtner_2013_Multiscale}
Fichtner, A., Trampert, J., Cupillard, P., Saygin, E., Taymaz, T., Capdeville, Y., and Villase{\~n}or, A.: Multiscale Full Waveform Inversion, Geophysical Journal International, 194, 534--556, \doi{10.1093/gji/ggt118}, 2013.

\bibitem[{Fichtner et~al.(2018)Fichtner, {van~Herwaarden}, Afanasiev, Simut{\.e}, Krischer, {\c C}ubuk-Sabuncu, Taymaz, Colli, Saygin, Villase{\~n}or, Trampert, Cupillard, Bunge, and Igel}]{fichtner_2018_Collaborative}
Fichtner, A., {van~Herwaarden}, D.-P., Afanasiev, M., Simut{\.e}, S., Krischer, L., {\c C}ubuk-Sabuncu, Y., Taymaz, T., Colli, L., Saygin, E., Villase{\~n}or, A., Trampert, J., Cupillard, P., Bunge, H.-P., and Igel, H.: The Collaborative Seismic Earth Model: Generation 1, Geophysical Research Letters, 45, 4007--4016, \doi{10.1029/2018gl077338}, 2018.

\bibitem[{Foti et~al.(2009)Foti, Comina, Boiero, and Socco}]{foti_2009_Nonuniqueness}
Foti, S., Comina, C., Boiero, D., and Socco, L.: Non-Uniqueness in Surface-Wave Inversion and Consequences on Seismic Site Response Analyses, Soil Dynamics and Earthquake Engineering, 29, 982--993, \doi{10.1016/j.soildyn.2008.11.004}, 2009.

\bibitem[{Foti et~al.(2014)Foti, Lai, Rix, and Strobbia}]{foti_2014_Surface}
Foti, S., Lai, C., Rix, G.~J., and Strobbia, C.: Surface Wave Methods for Near-Surface Site Characterization, CRC Press, 0 edn., ISBN 978-0-429-17853-5, \doi{10.1201/b17268}, 2014.

\bibitem[{Fu et~al.(2022)Fu, Pan, Li, Dong, Ma, and Chen}]{fu_2022_Improved}
Fu, L., Pan, L., Li, Z., Dong, S., Ma, Q., and Chen, X.: Improved High-resolution {{3D}} vs Model of Long Beach, {{CA}}: {{Inversion}} of Multimodal Dispersion Curves from Ambient Noise of a Dense Array, Geophysical Research Letters, 49, e2021GL097\,619, \doi{10.1029/2021GL097619}, 2022.

\bibitem[{Gan et~al.(2024)Gan, Yang, Pan, Sun, Zhang, Gao, and Chen}]{gan_2024_Deep}
Gan, Y., Yang, Z., Pan, L., Sun, Y.-C., Zhang, D., Gao, Y., and Chen, X.: Deep {{Learning-Based Dispersion Spectrum Inversion}} for {{Surface Wave Exploration}}, IEEE Transactions on Geoscience and Remote Sensing, 62, 1--11, \doi{10.1109/TGRS.2024.3399033}, 2024.

\bibitem[{Gao et~al.(2025)Gao, Wu, Sun, Hou, Gao, Wang, and Sheng}]{gao_2025_CigFacies}
Gao, H., Wu, X., Sun, X., Hou, M., Gao, H., Wang, G., and Sheng, H.: {{cigFacies}}: A Massive-Scale Benchmark Dataset of Seismic Facies and Its Application, Earth System Science Data, 17, 595--609, \doi{10.5194/essd-17-595-2025}, 2025.

\bibitem[{Haskell(1953)}]{haskell_1953_Dispersion}
Haskell, N.~A.: The Dispersion of Surface Waves on Multilayered Media, in: Vincit {{Veritas}}: {{A Portrait}} of the {{Life}} and {{Work}} of {{Norman Abraham Haskell}}, 1905--1970, edited by Ben-Menahem, A., vol.~43, pp. 86--103, American Geophysical Union, Washington, D. C., ISBN 978-0-87590-762-8, \doi{10.1785/BSSA0430010017}, 1953.

\bibitem[{Herrmann(2013)}]{herrmann_2013_Computer}
Herrmann, R.~B.: Computer Programs in Seismology: {{An}} Evolving Tool for Instruction and Research, Seismological Research Letters, 84, 1081--1088, \doi{10.1785/0220110096}, 2013.

\bibitem[{Ho et~al.(2020)Ho, Jain, and Abbeel}]{ho_2020_Denoising}
Ho, J., Jain, A., and Abbeel, P.: Denoising Diffusion Probabilistic Models, \doi{10.48550/arXiv.2006.11239}, 2020.

\bibitem[{Hu et~al.(2020)Hu, Qiu, Zhang, and {Ben-Zion}}]{hu_2020_Using}
Hu, J., Qiu, H., Zhang, H., and {Ben-Zion}, Y.: Using {{Deep Learning}} to {{Derive Shear-Wave Velocity Models}} from {{Surface-Wave Dispersion Data}}, Seismological Research Letters, 91, 1738--1751, \doi{10.1785/0220190222}, 2020.

\bibitem[{Huang et~al.(2024)Huang, Yu, Wang, and Wang}]{huang_2024_JointNet}
Huang, X., Yu, Z., Wang, W., and Wang, F.: {{JointNet}}: {{A Multimodal Deep Learning-Based Approach}} for {{Joint Inversion}} of {{Rayleigh Wave Dispersion}} and {{Ellipticity}}, Bulletin of the Seismological Society of America, 114, 627--641, \doi{10.1785/0120230199}, 2024.

\bibitem[{Jiang et~al.(2025)Jiang, Ma, Ning, Li, Wu, and Bao}]{jiang_2025_OneFitAll}
Jiang, Y., Ma, J., Ning, J., Li, J., Wu, H., and Bao, T.: One-{{Fit}}-{{All Transformer}} for {{Multimodal Geophysical Inversion}}: {{Method}} and {{Application}}, Journal of Geophysical Research: Machine Learning and Computation, 2, e2024JH000\,432, \doi{10.1029/2024JH000432}, 2025.

\bibitem[{Krischer et~al.(2018)Krischer, Fichtner, Boehm, and Igel}]{krischer_2018_Automated}
Krischer, L., Fichtner, A., Boehm, C., and Igel, H.: Automated Large-scale Full Seismic Waveform Inversion for North {{America}} and the North Atlantic, Journal of Geophysical Research: Solid Earth, 123, 5902--5928, \doi{10.1029/2017JB015289}, 2018.

\bibitem[{Lin et~al.(2013)Lin, Li, Clayton, and Hollis}]{lin_2013_Highresolution}
Lin, F.-C., Li, D., Clayton, R.~W., and Hollis, D.: High-Resolution {{3D}} Shallow Crustal Structure in Long Beach, California: {{Application}} of Ambient Noise Tomography on a Dense Seismic Array, Geophysics, 78, Q45--Q56, \doi{10.1190/geo2012-0453.1}, 2013.

\bibitem[{Liu et~al.(2024)Liu, Li, Fu, and Lu}]{liu_2024_Multimodal}
Liu, F., Li, J., Fu, L., and Lu, L.: Multimodal Surface Wave Inversion with Automatic Differentiation, Geophysical Journal International, 238, 290--312, \doi{10.1093/gji/ggae155}, 2024.

\bibitem[{Liu et~al.(2025)Liu, Deng, Su, Bai, and Ouyang}]{liu_2025_DispFormer}
Liu, F., Deng, B., Su, R., Bai, L., and Ouyang, W.: {{DispFormer}}: Pretrained Transformer for Flexible Dispersion Curve Inversion from Global Synthesis to Regional Applications, \doi{10.48550/ARXIV.2501.04366}, 2025.

\bibitem[{Lu et~al.(2018)Lu, Stehly, Paul, and {AlpArray Working Group}}]{lu_2018_Highresolution}
Lu, Y., Stehly, L., Paul, A., and {AlpArray Working Group}: High-Resolution Surface Wave Tomography of the European Crust and Uppermost Mantle from Ambient Seismic Noise, Geophysical Journal International, 214, 1136--1150, \doi{10.1093/gji/ggy188}, 2018.

\bibitem[{Luo et~al.(2022)Luo, Huang, Yang, Zhao, Yang, and Xu}]{luo_2022_Constructing}
Luo, Y., Huang, Y., Yang, Y., Zhao, K., Yang, X., and Xu, H.: Constructing Shear Velocity Models from Surface Wave Dispersion Curves Using Deep Learning, Journal of Applied Geophysics, 196, 104\,524, \doi{10.1016/j.jappgeo.2021.104524}, 2022.

\bibitem[{Merrifield et~al.(2022)Merrifield, Griffith, Zamanian, Gesbert, Sen, De~La Torre~Guzman, Potter, and Kuehl}]{merrifield_2022_Synthetic}
Merrifield, T.~P., Griffith, D.~P., Zamanian, S.~A., Gesbert, S., Sen, S., De~La Torre~Guzman, J., Potter, R.~D., and Kuehl, H.: Synthetic Seismic Data for Training Deep Learning Networks, Interpretation, 10, SE31--SE39, \doi{10.1190/INT-2021-0193.1}, 2022.

\bibitem[{Michelini et~al.(2021)Michelini, Cianetti, Gaviano, Giunchi, Jozinovi{\'c}, and Lauciani}]{michelini_2021_INSTANCE}
Michelini, A., Cianetti, S., Gaviano, S., Giunchi, C., Jozinovi{\'c}, D., and Lauciani, V.: {{INSTANCE}} -- the Italian Seismic Dataset for Machine Learning, Earth System Science Data, 13, 5509--5544, \doi{10.5194/essd-13-5509-2021}, 2021.

\bibitem[{Mousavi et~al.(2019)Mousavi, Sheng, Zhu, and Beroza}]{mousavi_2019_STanford}
Mousavi, S.~M., Sheng, Y., Zhu, W., and Beroza, G.~C.: {{STanford EArthquake}} Dataset ({{STEAD}}): {{A}} Global Data Set of Seismic Signals for {{AI}}, IEEE Access, 7, 179\,464--179\,476, \doi{10.1109/ACCESS.2019.2947848}, 2019.

\bibitem[{Park et~al.(1999)Park, Miller, and Xia}]{park_1999_Multichannel}
Park, C.~B., Miller, R.~D., and Xia, J.: Multichannel Analysis of Surface Waves, GEOPHYSICS, 64, 800--808, \doi{10.1190/1.1444590}, 1999.

\bibitem[{Pasyanos et~al.(2014)Pasyanos, Masters, Laske, and Ma}]{pasyanos_2014_LITHO10}
Pasyanos, M.~E., Masters, T.~G., Laske, G., and Ma, Z.: {{LITHO1}}.0: {{An}} Updated Crust and Lithospheric Model of the {{Earth}}, Journal of Geophysical Research: Solid Earth, 119, 2153--2173, \doi{10.1002/2013JB010626}, 2014.

\bibitem[{Reid et~al.(2025)Reid, Olivier, and Jones}]{reid_2025_Ambient}
Reid, A., Olivier, G., and Jones, T.: Ambient {{Noise Tomography}}: {{A Sensitive}}, {{Rapid}}, {{Passive Seismic Technique}} for {{Mineral Exploration}}, SEG Discovery, pp. 17--26, \doi{10.5382/SEGnews.2025-140.fea-01}, 2025.

\bibitem[{Rickers et~al.(2013)Rickers, Fichtner, and Trampert}]{rickers_2013_Iceland}
Rickers, F., Fichtner, A., and Trampert, J.: The {{Iceland}}--Jan Mayen Plume System and Its Impact on Mantle Dynamics in the North Atlantic Region: Evidence from Full-Waveform Inversion, Earth and Planetary Science Letters, 367, 39--51, \doi{10.1016/j.epsl.2013.02.022}, 2013.

\bibitem[{Shapiro and Campillo(2004)}]{shapiro_2004_Emergence}
Shapiro, N.~M. and Campillo, M.: Emergence of Broadband {{Rayleigh}} Waves from Correlations of the Ambient Seismic Noise, Geophysical Research Letters, 31, 2004GL019\,491, \doi{10.1029/2004GL019491}, 2004.

\bibitem[{Shapiro and Ritzwoller(2002)}]{shapiro_2002_MonteCarlo}
Shapiro, N.~M. and Ritzwoller, M.~H.: Monte-{{Carlo}} Inversion for a Global Shear-Velocity Model of the Crust and Upper Mantle, Geophysical Journal International, 151, 88--105, \doi{10.1046/j.1365-246X.2002.01742.x}, 2002.

\bibitem[{Shen et~al.(2013)Shen, Ritzwoller, and Schulte-Pelkum}]{shen_2013_3D}
Shen, W., Ritzwoller, M.~H., and Schulte-Pelkum, V.: A 3-{{D}} Model of the Crust and Uppermost Mantle beneath the {{Central}} and {{Western US}} by Joint Inversion of Receiver Functions and Surface Wave Dispersion, Journal of Geophysical Research: Solid Earth, 118, 262--276, \doi{10.1029/2012JB009602}, 2013.

\bibitem[{Shen et~al.(2016)Shen, Ritzwoller, Kang, Kim, Lin, Ning, Wang, Zheng, and Zhou}]{shen_2016_Seismic}
Shen, W., Ritzwoller, M.~H., Kang, D., Kim, Y., Lin, F.-C., Ning, J., Wang, W., Zheng, Y., and Zhou, L.: A Seismic Reference Model for the Crust and Uppermost Mantle beneath {{China}} from Surface Wave Dispersion, Geophysical Journal International, 206, 954--979, \doi{10.1093/gji/ggw175}, 2016.

\bibitem[{Simut{\.e} et~al.(2016)Simut{\.e}, Steptoe, Cobden, Gokhberg, and Fichtner}]{simutė_2016_Fullwaveform}
Simut{\.e}, S., Steptoe, H., Cobden, L., Gokhberg, A., and Fichtner, A.: Full-waveform Inversion of the Japanese Islands Region, Journal of Geophysical Research: Solid Earth, 121, 3722--3741, \doi{10.1002/2016jb012802}, 2016.

\bibitem[{Socco and Strobbia(2004)}]{socco_2004_Surfacewave}
Socco, L. and Strobbia, C.: Surface-wave Method for Near-surface Characterization: A Tutorial, Near Surface Geophysics, 2, 165--185, \doi{10.3997/1873-0604.2004015}, 2004.

\bibitem[{Taufik et~al.(2024)Taufik, Wang, and Alkhalifah}]{taufik_2024_Learned}
Taufik, M.~H., Wang, F., and Alkhalifah, T.: Learned {{Regularizations}} for {{Multi}}-{{Parameter Elastic Full Waveform Inversion Using Diffusion Models}}, Journal of Geophysical Research: Machine Learning and Computation, 1, e2024JH000\,125, \doi{10.1029/2024JH000125}, 2024.

\bibitem[{Thomson(1950)}]{thomson_1950_Transmission}
Thomson, W.~T.: Transmission of Elastic Waves through a Stratified Solid Medium, Journal of Applied Physics, 21, 89--93, \doi{10.1063/1.1699629}, 1950.

\bibitem[{Wang et~al.(2022)Wang, Song, and Li}]{wang_2022_Deep}
Wang, F., Song, X., and Li, J.: Deep {{Learning}}-{{Based}} {{{\emph{H}}}}{\emph{-{$\kappa$}}} {{Method}} ({{HkNet}}) for {{Estimating Crustal Thickness}} and {{{\emph{Vp}}}}{\emph{/}}{{{\emph{Vs}}}} {{Ratio From Receiver Functions}}, Journal of Geophysical Research: Solid Earth, 127, e2022JB023\,944, \doi{10.1029/2022JB023944}, 2022.

\bibitem[{Wang et~al.(2023{\natexlab{a}})Wang, Huang, and Alkhalifah}]{wang_2023_Prior}
Wang, F., Huang, X., and Alkhalifah, T.~A.: A {{Prior Regularized Full Waveform Inversion Using Generative Diffusion Models}}, IEEE Transactions on Geoscience and Remote Sensing, 61, 1--11, \doi{10.1109/tgrs.2023.3337014}, 2023{\natexlab{a}}.

\bibitem[{Wang et~al.(2023{\natexlab{b}})Wang, Song, and Li}]{wang_2023_Deeplearningbased}
Wang, F., Song, X., and Li, M.: A Deep-Learning-Based Approach for Seismic Surface-Wave Dispersion Inversion ({{SfNet}}) with Application to the {{Chinese}} Mainland, Earthquake Science, 36, 147--168, \doi{10.1016/j.eqs.2023.02.007}, 2023{\natexlab{b}}.

\bibitem[{Wang et~al.(2025)Wang, Wu, and Zhang}]{wang_2025_CigChannel}
Wang, G., Wu, X., and Zhang, W.: {{cigChannel}}: A Large-Scale {{3D}} Seismic Dataset with Labeled Paleochannels for Advancing Deep Learning in Seismic Interpretation, Earth System Science Data, 17, 3447--3471, \doi{10.5194/essd-17-3447-2025}, 2025.

\bibitem[{Wathelet et~al.(2004)Wathelet, Jongmans, and Ohrnberger}]{wathelet_2004_Surfacewave}
Wathelet, M., Jongmans, D., and Ohrnberger, M.: Surface-wave Inversion Using a Direct Search Algorithm and Its Application to Ambient Vibration Measurements, Near Surface Geophysics, 2, 211--221, \doi{10.3997/1873-0604.2004018}, 2004.

\bibitem[{Wen et~al.(2023)Wen, Yu, {Department of Geosciences, State University of New York at Stony Brook, Stony Brook, NY 11794, USA}, {Laboratory of Seismology and Physics of Earth's Interior; School of Earth and Space Sciences, University of Science and Technology of China, Hefei 230026, China}, and {Department of Earth Sciences, National Natural Science Foundation of China, Beijing 100085, China}}]{wen_2023_China}
Wen, L., Yu, S., {Department of Geosciences, State University of New York at Stony Brook, Stony Brook, NY 11794, USA}, {Laboratory of Seismology and Physics of Earth's Interior; School of Earth and Space Sciences, University of Science and Technology of China, Hefei 230026, China}, and {Department of Earth Sciences, National Natural Science Foundation of China, Beijing 100085, China}: The {{China}} Seismological Reference Model Project, Earth and Planetary Physics, 7, 521--532, \doi{10.26464/epp2023078}, 2023.

\bibitem[{Xia et~al.(1999)Xia, Miller, and Park}]{xia_1999_Estimation}
Xia, J., Miller, R.~D., and Park, C.~B.: Estimation of Near-surface Shear-wave Velocity by Inversion of {{Rayleigh}} Waves, GEOPHYSICS, 64, 691--700, \doi{10.1190/1.1444578}, 1999.

\bibitem[{Xiao et~al.(2024)Xiao, Cheng, Wu, Wang, Sun, Wang, Ma, Tong, Liang, Tian, Li, Chen, Yu, and Wen}]{xiao_2024_CSRM10}
Xiao, X., Cheng, S., Wu, J., Wang, W., Sun, L., Wang, X., Ma, J., Tong, Y., Liang, X., Tian, X., Li, H., Chen, Q.-F., Yu, S., and Wen, L.: {{CSRM}}-1.0: {{A China Seismological Reference Model}}, Journal of Geophysical Research: Solid Earth, 129, e2024JB029\,520, \doi{10.1029/2024JB029520}, 2024.

\bibitem[{Xie et~al.(2018)Xie, Chu, and Yang}]{xie_2018_3D}
Xie, J., Chu, R., and Yang, Y.: 3-{{D}} Upper-Mantle Shear Velocity Model beneath the Contiguous United States Based on Broadband Surface Wave from Ambient Seismic Noise, Pure and Applied Geophysics, 175, 3403--3418, \doi{10.1007/s00024-018-1881-2}, 2018.

\bibitem[{Xin et~al.(2019)Xin, Zhang, Kang, He, Gao, and Gao}]{xin_2019_Highresolution}
Xin, H., Zhang, H., Kang, M., He, R., Gao, L., and Gao, J.: High-resolution Lithospheric Velocity Structure of Continental China by Double-difference Seismic Travel-time Tomography, Seismological Research Letters, 90, 229--241, \doi{10.1785/0220180209}, 2019.

\bibitem[{Yablokov et~al.(2023)Yablokov, Lugovtsova, and Serdyukov}]{yablokov_2023_Uncertainty}
Yablokov, A., Lugovtsova, Y., and Serdyukov, A.: Uncertainty Quantification of Multimodal Surface Wave Inversion Using Artificial Neural Networks, GEOPHYSICS, 88, KS1--KS11, \doi{10.1190/geo2022-0261.1}, 2023.

\bibitem[{Yablokov et~al.(2021)Yablokov, Serdyukov, Loginov, and Baranov}]{yablokov_2021_Artificial}
Yablokov, A.~V., Serdyukov, A.~S., Loginov, G.~N., and Baranov, V.~D.: An Artificial Neural Network Approach for the Inversion of Surface Wave Dispersion Curves, Geophysical Prospecting, 69, 1405--1432, \doi{10.1111/1365-2478.13107}, 2021.

\bibitem[{Yang and Ritzwoller(2008)}]{yang_2008_Characteristics}
Yang, Y. and Ritzwoller, M.~H.: Characteristics of Ambient Seismic Noise as a Source for Surface Wave Tomography, Geochemistry, Geophysics, Geosystems, 9, 2007GC001\,814, \doi{10.1029/2007GC001814}, 2008.

\end{thebibliography}

\end{document}